\documentclass[aps,prb,pdflatex,reprint,showpacs,superscriptaddress]{revtex4-2}
\usepackage{amssymb,bbold}
\usepackage{amsmath}
\usepackage{dsfont}
\usepackage[utf8x]{inputenc}
\usepackage[usenames,dvipsnames]{xcolor}
\usepackage{natbib}
\usepackage{hyperref}
\usepackage{epstopdf}
\usepackage{soul}
\usepackage{physics}
\usepackage{float}
\usepackage{amsfonts}
\usepackage[font=normal,labelfont=bf, justification=Justified]{caption}
\usepackage{lipsum}
\usepackage{subcaption}
\usepackage{mathtools}
\usepackage[normalem]{ulem}
\usepackage{braket}
\usepackage{booktabs}
\usepackage{multirow}
\hypersetup{
  colorlinks=true,linkcolor=blue,citecolor=blue,
  filecolor=blue,urlcolor=blue,breaklinks=true
}

\begin{document}
\title{Magnetoresistance and electric current oscillations induced by geometry \\in a two-dimensional quantum ring}
\author{Francisco A. G. de Lira}
\email{goncalves.francisco@discente.ufma.br \\  francisco.augoncalves@gmail.com}
\affiliation{
        Departamento de F\'{i}sica,
        Universidade Federal do Maranh\~{a}o,
        65085-580, S\~{a}o Lu\'{i}s, Maranh\~{a}o, Brazil
      }

\author{Edilberto O. Silva}
\email{edilberto.silva@ufma.br}
\affiliation{
Departamento de F\'{i}sica,
Universidade Federal do Maranh\~{a}o,
65085-580, S\~{a}o Lu\'{i}s, Maranh\~{a}o, Brazil
}

\author{Christian D. Santangelo}
\email{cdsantan@syr.edu}
\affiliation{Physics Department, Syracuse University, 13244 Syracuse, New York, United States}

\date{\today }

\begin{abstract}

In this work, we investigate the effects of a controlled conical geometry on the electric charge transport through a two-dimensional quantum ring weakly coupled to both the emitter and the collector. These mesoscopic systems are known for being able to confine highly mobile electrons in a defined region of matter. In particular, we consider a GaAs device having an average radius of $800\hspace{0.05cm}\text{nm}$ in different regimes of subband occupation at non-zero temperature and under the influence of a weak and uniform background magnetic field. Using the adapted Landauer formula for the resonant tunneling and the energy eigenvalues, we explore how the modified surface affects the Van-Hove conductance singularities, the magnetoresistance interference patterns resulting from the Aharonov-Bohm oscillations of different frequencies and the charge transport when an electric potential is applied in the terminals of the device. Magnetoresistance and charge current oscillations depending only on the curvature intensity are reported, providing a new feature that represents an alternative way to optimize the transport through the device by tuning its geometry.   
\end{abstract}

\maketitle
		
\section{Introduction}\label{intro}

The surge of interest in studying physical phenomena arising from quantum loop systems emerged due to the noted effect proposed by Y. Aharonov and D. Bohm in 1959 \cite{PR.1959.115.485}. Later, upon the development of the first Metal-Oxide Semiconductor Field Transistor (MOSFET) in 1960 \cite{KahngAtalla}, the prediction of integer quantum Hall effect quantization in 1975 \cite{doi:10.1143/JPSJ.39.279} and its first observation in 1980 \cite{PhysRevLett.45.494}, R. B. Laughlin proposed a thought experiment using a metallic loop in 1981 to show that quantized Hall conductivity is a consequence of gauge invariance \cite{PhysRevB.23.5632}. In the early '90s, M. B\"uttiker, Y. Imry, and R. Landauer predicted the existence of persistent current in non-superconducting metal loops \cite{BUTTIKER1983365, PhysRevLett.54.2049}, which was experimentally observed years later \cite{PhysRevLett.64.2074, PhysRevLett.67.3578}. Furthermore, in 1993, Maily et al. successfully performed these experiments in semiconductor rings \cite{PhysRevLett.70.2020}.

In recent years, experiments \cite{PhysRevB.77.085413, PhysRevB.78.193405, https://doi.org/10.1002/pssb.200982284} and theoretical models \cite{SST.1996.11.1635, PhysRevB.73.235327} have explored and predicted important aspects of these systems in more detail, including their electronic, magnetic, thermodynamic, transport, optical, and quantum informational properties \cite{PhysRevB.60.5626, LAFAURIE2024415786, LIMA2023169547, 10.1063/1.3688037, e24081059}. These studies have been leading the development of several applications in technology, such as THz detectors \cite{10.1063/1.3100407}, improved solar cell efficiency \cite{Wu2013-vk}, memory devices \cite{10.1063/1.3688037}, single-photon emitters \cite{Warburton2000-tg, PhysRevB.79.085308} and qubits for quantum computing \cite{Zipper_2011, PhysRevB.78.235312}.

Although this field is very diverse, a considerable branch of these investigations has recently turned to understand how geometry affects the properties of quantum rings and other structures able to confine electrons \cite{PhysRevB.69.195313, 10.1063/1.2206135, PhysRevB.75.205309, PhysRevLett.100.230403, PhysRevB.97.241103, PEREIRA2021114760, RIBEIRO2020100045} due to the existence of a purely geometrical potential arising from curvature as a result of the procedure implemented by R. T. C. da Costa \cite{PRA.1981.23.1982}. This is, indeed, an important concern due to the presence of topological defects in solid and liquid crystals, such as disclinations and screw dislocations \cite{Lin2023-le, GUTKIN2011329, KLEMAN2016}. Also, materials engineering has already made it possible to develop two-dimensional lattices with different geometries by modifying part of the structure \cite{PhysRevLett.115.026801, doi:10.1063/1.1377859}.  

Against this backdrop, conductance plays a key role since it is a quantity of great experimental interest \cite{PRB.1993.48.15148, PhysRevB.109.245405, PhysRevB.109.235107}. Thus, in this work, we investigate the effects of a disclination, described by the geometry of a cone, on the electric charge transport properties. In particular, we consider a two-dimensional quantum ring of GaAs with an average radius of $800 \hspace{0.05cm}\text{nm}$ at the temperature of $40 \hspace{0.05cm}\text{mK}$ and under the influence of a weak background magnetic field. We explore two regimes of fixed electron occupation: one ($0.5\hspace{0.05 cm}\text{meV}$) and four occupied subbands ($2\hspace{0.05 cm}\text{meV}$) \cite{PhysRevB.53.6947}.

This paper is outlined as follows. In Sec. \ref{qrgeometry}, we describe the model and the main aspects of differential geometry to introduce the concept of geometric potential. In Sec. \ref{landauer}, the basic concepts of charge transport, such as quantum of conductance, resonant tunneling, and coherent and incoherent transmission, are presented. In Sec. \ref{result}, we show and discuss the effect of the modified surface in the conductance in terms of the Fermi energy, the behavior of the total and average magnetoresistance with the magnetic field for different regimes of curved surface, highlighting an interesting and almost periodic dependence with the curvature parameter. Later, we investigate how the electric current behaves when an electric potential is applied through the device and its dependence with the geometry. Finally, we summarize the results in Sec. \ref{conclusions}.  

\section{Two-dimensional quantum ring with conical geometry}\label{qrgeometry}

\begin{figure}[t]
\centering
\includegraphics[width=0.5\textwidth]{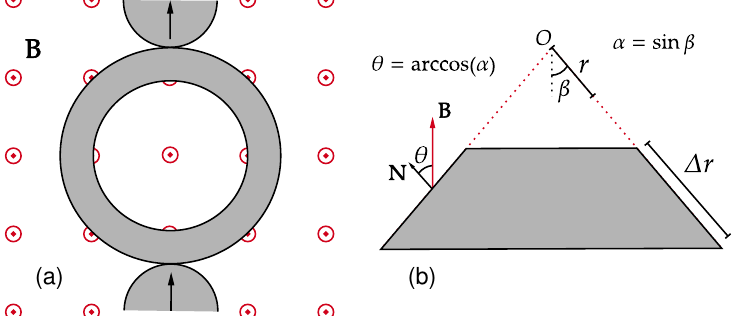}
\caption{(a) (Color online) Schematic illustration in an upper view of the weakly coupled ring model used in this work, adapted from \cite{PhysRevB.53.6947}. (b) A side view of the model highlights the conical geometry of the ring surface for $\alpha<1$. The origin is set on the virtual apex of the conical surface defined by the ring.}
\label{ringdevice}	
\end{figure}

The system of interest is a two-dimensional quantum ring weakly coupled to two leads working as a source and a drain, also called emitter and collector. The device is located in a region with a weak background magnetic field, as illustrated in Fig. \ref{ringdevice}a. The problem can be described using the theoretical model proposed by W. Tan and J. Inkson \cite{SST.1996.11.1635}, which provides a radial confining potential for the two-dimensional electron gas (2DEG) depending on two parameters, $a_{1}$ and $a_{2}$, in the form
\begin{equation}
V(r)=\dfrac{a_{1}}{r^{2}}+a_{2}r^{2}-2\sqrt{a_{1}a_{2}}.
\label{ring01}
\end{equation}
These parameters can be measured experimentally and be suitably controlled to describe other similar systems, such as the quantum dot ($a_{1}=0)$ and the antidot ($a_{2}=0$). This potential has a minimum at $r_{0}=\left(a_{1}/a_{2}\right)^{1/4}$, the average radius of the ring. Near this point, we can approximate expression (\ref{ring01}) as a parabola, $V(r)\approx \mu\omega^{2}_{0}(r-r_{0})^{2}/2$, from which we can define $\omega_{0}=\sqrt{8a_{2}/\mu}$ as a frequency related to the strength of the confinement where $\mu$ is the electron effective mass. Moreover, if the system has a total Fermi energy $\epsilon_{f}$, we can define an effective width, 
\begin{equation}
\Delta r=\sqrt{\dfrac{8\epsilon_{f}}{\mu\omega^{2}_{0}}}.
\label{ring012}
\end{equation}

We are interested in investigating the effects of curvature in the surface defined by this two-dimensional quantum ring. More precisely, we are concerned with the conical geometry described by the parametric equation (See Fig.\ref{ringdevice}b)
\begin{equation}
\mathbf{x}(r,\theta)=(\alpha r\cos\theta, \alpha r\sin\theta, -r\sqrt{1-\alpha^2}),
\label{ring02}
\end{equation}
where $r\in\mathbb{R^{+}}$ and $0\le\theta\le 2\pi$ are the surface parameters and $0<\alpha\le 1$ is the curvature parameter. The line element is obtained from the coefficients of the first fundamental form $E=\mathbf{x}_{rr}\cdot\mathbf{x}_{rr}=1$, $F=\mathbf{x}_{r\varphi}\cdot\mathbf{x}_{r\varphi}=0$ and $G=\mathbf{x}_{\varphi\varphi}\cdot\mathbf{x}_{\varphi\varphi}=\alpha^{2}r^{2}$, as
\begin{equation}
ds^{2}=g_{ij}x^{i}x^{j}=dr^{2}+\alpha^{2}r^{2}d\varphi^{2}.
\label{ring03}
\end{equation}
If we consider a magnetic field in $z$ direction, the unit normal vector of the surface
\begin{equation}
\mathbf{N}=(\sqrt{1-\alpha^{2}}\cos\varphi, \sqrt{1-\alpha^{2}}\sin\varphi, \alpha)
\label{ring032}
\end{equation}
is no longer parallel to $\mathbf{B}$ if $\alpha\neq 1$. Thus, the corresponding vector potential will be \cite{EPJP.2014.129.100}
\begin{equation}
\mathbf{A}=\dfrac{B}{2}\hat{e}_{\varphi}=\dfrac{B}{2}\alpha r\hat{\varphi}.
\label{ring04}
\end{equation}
Note that the curl of the vector potential (\ref{ring04}),  for which $\mathbf{N}$ is the orthonormal basis vector instead of $\hat{z}$, is equal to $B\alpha \mathbf{N}$. This is just the component of $B$ in the normal direction since $\alpha=\cos\theta$ (Fig. \ref{ringdevice}b).

The R. T. C da Costa procedure \cite{PRA.1981.23.1982} of confining a quantum particle to a surface tells us that this particle experiences a purely geometric potential arising from the curvature of the surface. With the coefficients of the second fundamental form $e=\mathbf{x}_{rr}\cdot\mathbf{N}=0$, $f=\mathbf{x}_{r\varphi}\cdot\mathbf{N}=0$ and $g=\mathbf{x}_{\varphi\varphi}\cdot\mathbf{N}=-\alpha r\sqrt{1-\alpha^{2}}$, we can explicitly find an expression for the geometric potential
\begin{equation}
V_{S}=-\dfrac{\hbar^{2}}{2\mu}\left(M^{2}-K\right),
\label{ring05}
\end{equation}
where 
\begin{equation}
M=\dfrac{Eg-2Ff+Ge}{2(EG-F^2)}\quad\text{and}\quad  K=\dfrac{eg-f^{2}}{EG-F^2}
\label{ring06}
\end{equation}
are the mean and Gaussian curvatures, respectively. Thus, the geometric potential for the parametric surface in (\ref{ring02}) is given by the expression \cite{Carvalho_2007}
\begin{equation}
V_{S}=-\dfrac{\hbar^{2}}{2\mu}\left[\dfrac{(1-\alpha^{2})}{4\alpha^{2}r^{2}}-\left(\dfrac{1-\alpha}{\alpha}\right)\dfrac{\delta(r)}{r}\right].
\label{ring07}
\end{equation}
We can clearly see that this potential provides an attractive force towards the center of the ring (conic apex), whereas an extremely repulsive contribution exists at its very center.

The corresponding two-dimensional Schr\"odinger equation of the system can now be computed in its general form as
\begin{align}
E\chi_{S}=&-\dfrac{\hbar^{2}}{2\mu}\left[\dfrac{1}{\sqrt{g}}\dfrac{\partial}{\partial q^{i}}\left(\sqrt{g}g^{ij}\dfrac{\partial\chi_{S}}{\partial q^{j}}\right)\right]\notag\\&-\dfrac{\hbar^{2}}{2\mu}\left[\dfrac{ie}{\hbar\sqrt{g}}\dfrac{\partial }{\partial q^{i}}\left(\sqrt{g}g^{ij} A_{j}\right)\chi_{S}\right]\notag\\
&-\dfrac{\hbar^{2}}{2\mu}\left[\dfrac{2ie}{\hbar}g^{ij}A_{i}\dfrac{\partial\chi_{S}}{\partial q^{j}}-\dfrac{e^{2}}{\hbar^{2}}g^{ij}A_{i}A_{j}\chi_{S}\right]\notag\\
&-V_{S}\chi_{S} +\left(\dfrac{a_{1}}{r^{2}}+a_{2}r^{2}-V_{0}\right)\chi_{S},
\label{ring08}
\end{align}
where $\chi_{S}=\chi_{S}(r,\varphi)$ represents the surface wavefunction resulting from the da Costa procedure \cite{PRA.1981.23.1982} and $i,j=\{1,2\}$. Using result (\ref{ring03}) and (\ref{ring07}) in the last expression, we obtain the explicit equation for the conical geometry
\begin{align}
-&\dfrac{\hbar^{2}}{2\mu}\left[\dfrac{1}{r}\dfrac{\partial}{\partial r}\left(r\dfrac{\partial}{\partial r}\right)-\dfrac{e^{2}B^{2}\alpha^{2}r^{2}}{4\hbar^{2}}+\dfrac{1}{\alpha^{2}r^{2}}\left(\dfrac{\partial}{\partial \varphi}+il\right)^{2}\right]\chi_{S}\notag \\-&\dfrac{\hbar^{2}}{2\mu}\dfrac{ieB}{\hbar}\left(\dfrac{\partial}{\partial \varphi}+il\right)-\dfrac{\hbar^{2}}{2\mu}\left[\dfrac{(1-\alpha^{2})}{4\alpha^{2}r^{2}}-\left(\dfrac{1-\alpha}{\alpha}\right)\dfrac{\delta(r)}{r}\right]\chi_{S}\notag \\
+&\left(\dfrac{a_{1}}{r^{2}}+a_{2}r^{2}-V_{0}\right)\chi_{S}=E\chi_{S}.
\label{ring09}
\end{align}
Proposing solutions of the form $\chi_{S}(r,\varphi)=e^{im\varphi}f(r)$ in the previous expression, we obtain the radial differential equation
\begin{align}
&\dfrac{d^{2}f}{dr^{2}}+\dfrac{1}{r}\dfrac{df}{dr}-\left[\dfrac{2\mu a_{2}}{\hbar^{2}}+\dfrac{e^{2}B^{2}\alpha^{2}}{4\hbar^{2}}\right]r^{2}f(r)\notag\\
&-\dfrac{1}{r^{2}}\left[\dfrac{2\mu a_{1}}{\hbar^{2}}+\dfrac{(m-l)^{2}}{\alpha^{2}}-\dfrac{(1-\alpha^{2})}{4\alpha^{2}}\right]f(r)\notag\\
&-\left\{\left[\dfrac{eB(m-l)}{\hbar}+\dfrac{2\mu(E+V_{0})}{\hbar^2}\right]-\dfrac{(1-\alpha)}{\alpha r}\delta(r)\right\}f(r)=0,
\label{ring10}
\end{align}
from which is useful to define
\begin{equation}
\dfrac{2\mu a_{2}}{\hbar^{2}}+\dfrac{e^{2}B^{2}\alpha^{2}}{4\hbar^{2}}=\dfrac{1}{4\lambda^{4}}\quad\text{and}\quad\omega=\sqrt{\omega^{2}_{0}+(\alpha\omega_{c})^{2}},\label{ring11}
\end{equation}
as the effective magnetic length $\lambda=\sqrt{\hbar/\mu\omega}$ and $\omega_{c}=eB/\mu$ as the cyclotron frequency. Moreover, we can rewrite
\begin{equation} 
L=\sqrt{\dfrac{2\mu a_{1}}{\hbar^{2}}+\dfrac{m^{2}}{\alpha^{2}}-\dfrac{(1-\alpha^{2})}{4\alpha^{2}}}\label{ring12}
\end{equation}
and 
\begin{equation}
S=\sqrt{\dfrac{eB m}{\hbar}+\dfrac{2\mu(E+V_{0})}{\hbar^2}},\label{ring13}
\end{equation}
in order to remain with a compact form for the radial differential equation (\ref{ring10}) \cite{PEREIRA2021114760}
\begin{equation}
\dfrac{d^{2}f}{dr^{2}}+\dfrac{1}{r}\dfrac{df}{dr}-\left[\dfrac{r^{2}}{4\lambda^{4}}+\dfrac{L^{2}}{r^{2}}-S^{2}+\dfrac{(1-\alpha)}{\alpha r}\delta(r)\right]f(r)=0.
\label{ring14}
\end{equation}

\begin{figure}[t!]
\centering
\begin{subfigure}{0.5\textwidth}
\includegraphics[width=\textwidth]{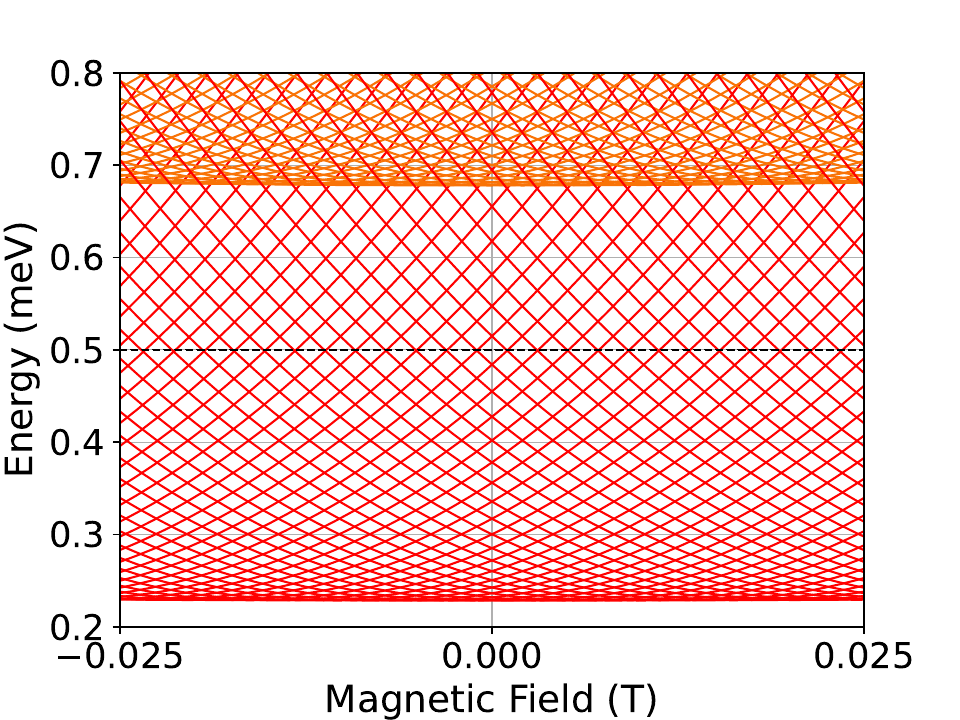}
\subcaption{$\alpha=1$}
\label{subband1}
\end{subfigure}
\begin{subfigure}{0.5\textwidth}
\includegraphics[width=\textwidth]{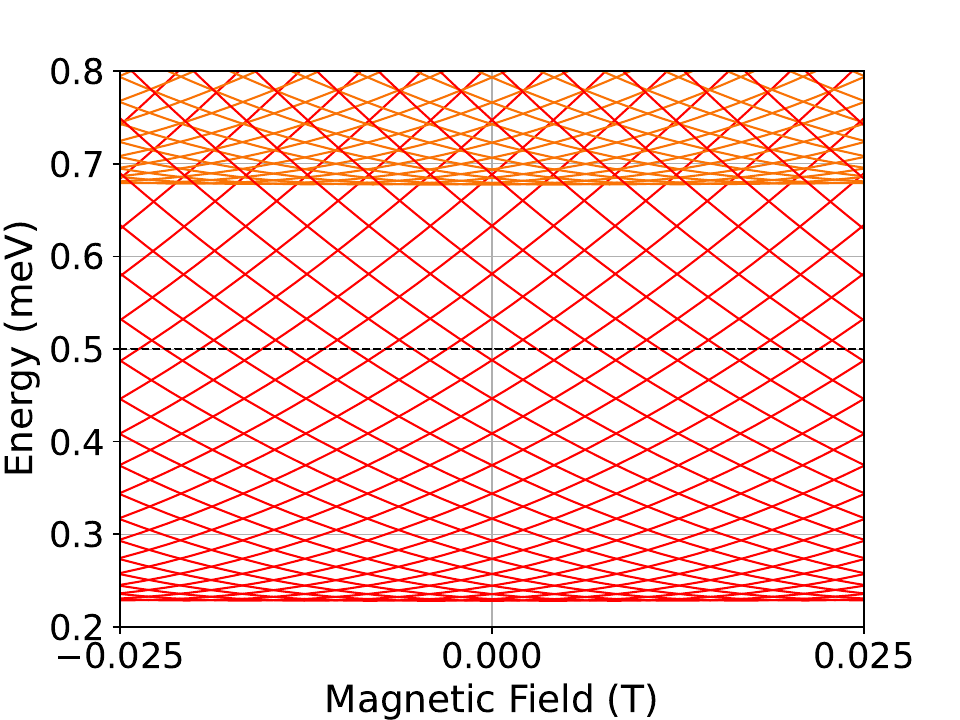}
\subcaption{$\alpha=0.7$}
\label{subband07}
\end{subfigure}
\caption{(Color online) Energy eigenvalues of the quantum ring near $\epsilon_{f}=0.5\hspace{0.05cm}\text{meV}$ as a function of the magnetic field $B$ for flat (a) and curvature (b) cases. Each curve denotes a specific $\chi_{n,m}$ state, and each color represents different subbands. In Figs. \ref{subband1} and \ref{subband07}, we plotted the subbands $n=0$ (red) and $n=1$ (orange). The dashed line locates the Fermi level.}
\label{subbands}
\end{figure}

As discussed in \cite{AndradeSilvaPereira2012}, due to the presence of the $\delta$ function in the Hamiltonian operator in Eq. (\ref{ring09}), $\hat{H}$ is not self-adjoint in general, and singular solutions must be obtained through the self-adjoint extension method \cite{PRL.1990.64.503, AndradeSilvaPereira2012}. However, it can be shown that the Hamiltonian is essentially self-adjoint if $\abs{L}\ge 1$, and thus, we can neglect the influence of the $\delta$ function and consider only the regular solutions \cite{AndradeSilvaPereira2012, Book.1988.Albeverio} in Eq. (\ref{ring14}). In fact, for a quantum ring in the mesoscopic regime, we can assert that $L>1$ and, consequently, the eigenfunctions and eigenvalues of the system are given by \cite{DELIRA2024115898, PEREIRA2021114760}:
\begin{align}
\chi_{n,m}(r,\varphi)&=\dfrac{1}{\lambda}\sqrt{\dfrac{\Gamma(L+n+1)}{2^{L+1}\pi \left[\Gamma(L+1)\right]^2 n!}}\left(\frac{r}{\lambda}\right)^{L}\notag\\&\times e^{-\frac{1}{4}\left(\frac{r}{\lambda}\right)^2}\prescript{}{1}{F}_{1}\left(-n, L+1, \dfrac{1}{2}\left(\frac{r}{\lambda}\right)^2\right)e^{im\varphi},
\label{ring15}
\end{align}
\begin{equation}
E_{n,m}=\left(n+\dfrac{1}{2}+\dfrac{L}{2}\right)\hbar\omega-\dfrac{m}{2}\hbar\omega_{c}-\dfrac{\mu}{4}\omega_{0}^{2}r^{2}_{0},
\label{ring16}
\end{equation}
where $_{1}F_{1}$ is the confluent hypergeometric function of the first kind, $n=\{0,1,2,3,\dots\}$  is the radial quantum number which denotes the subband level, and $m=\{\dots -2, -1, 0, 1, 2, \dots\}$ is the orbital quantum number which specifies the position in the subband. 
Note that $\alpha$ directly modulates the influence of the magnetic field since the flux of $B$ through the surface enclosed by the ring reduces as $\alpha$ decreases. For $\alpha=1$, we recover previously reported results \cite{PRB.1999.60.5626}.

In Figs. \ref{subband1} and \ref{subband07}, we plot energy levels of the system as a function of the magnetic field for the cases of $\alpha=1$ (flat) and $\alpha=0.7$, respectively. Each curve matches a specific value of both quantum numbers $n$ and $m$, where we are using different colors to represent the levels $n=0$ (red) and $n=1$ (orange). Note that the border between regions with different colors outlines the behavior of the subband minimum with $B$. 

\begin{figure}[b!]
\centering
\begin{subfigure}{0.5\textwidth}
\includegraphics[width=\textwidth]{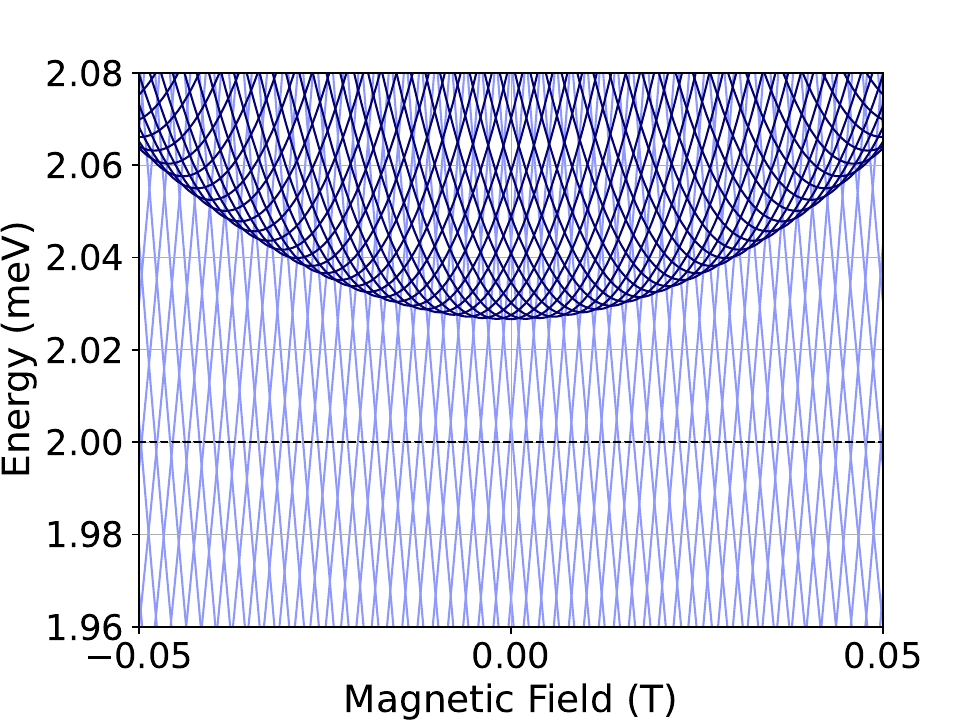}
\subcaption{$\alpha=1$}
\label{subbandsmin}
\end{subfigure}
\begin{subfigure}{0.5\textwidth}
\includegraphics[width=\textwidth]{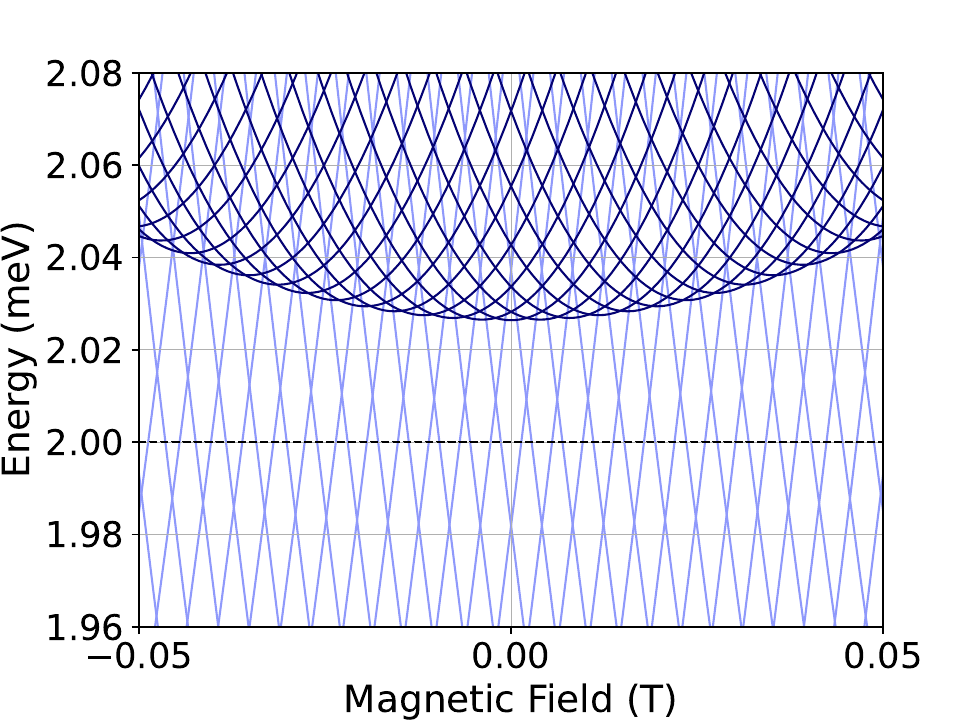}
\subcaption{$\alpha=0.7$}
\label{subbandsmin07}
\end{subfigure}
\caption{(Color online) Energy eigenvalues of the quantum ring near $\epsilon_{f}=2\hspace{0.05cm}\text{meV}$ as a function of the magnetic field $B$ for flat (a) and curvature (b) cases. We plotted the subbands $n=3$ (light blue) and $n=4$ (dark blue). We have omitted the states from the lower subbands ($n\le 2$) for better visualization.}
\label{subbandscurvature}
\end{figure}

As the magnetic field varies at $0\hspace{0.05cm}\text{K}$, electrons in the ring perform transitions upon reaching a state crossing point to maintain their most stable configuration. Being $r_{n,m}=\sqrt{2L}\lambda$ the average radius of a state, these transitions generate an oscillation in their energies, known as Aharonov-Bohm (AB) oscillations, with period $p_{0}=\phi_{0}/\pi r^{2}_{n,m}\approx 2.037\hspace{0.05cm}\text{mT}$ around $0.5\hspace{0.05cm}\text{meV}$ and $\phi_{0}=h/e$ being the quantum of flux \cite{PhysRevB.53.6947}. One can identify that, as the curvature in the surface increases, the period of the AB oscillations rises by $p_{0}/\alpha^{2}$ \cite{PhysRevB.53.6947, PEREIRA2021114760, DELIRA2024115898} as a consequence of its dependence with the area $\pi(\alpha r_{n,m})^{2}$ enclosed by the loop. On the other hand, the density of states per unit of energy reduces as a result of the growth in the gap between energetic neighbors already  observed in 
\cite{DELIRA2024115898}.

In Figs \ref{subbandsmin} and \ref{subbandsmin07}, we perform the same analysis for a regime of four occupied subbands, where states near $\epsilon_{f}=2\hspace{0.05cm} \text{meV}$ from $n=3$ (light blue) and $n=4$ (dark blue) are highlighted, omitting those from the lower levels to obtain a better visualization of the energy spectrum. A comparison between the flat and curvature cases shows us the same aspects discussed before about the AB period and density of states. However, this situation also provides us with another effect of the modified surface, which can be identified by a reduction in the increment rate of subband minimum $n=4$ with the strength of the magnetic field. This is a consequence of attenuation in the influence of $\mathbf{B}$ from the conical geometry, as expected in Eq. (\ref{ring11}). Indeed, we can see no significant change when $B=0$. 

As we will explore in the next sections, the three factors mentioned above: AB periods, density of states per unity of energy and subband minimum are going to have important functions in the manifestation of the properties we are investigating. 

\section{Landauer formula and resonant tunneling on weakly coupled ring}\label{landauer}

Classically, one should expect the conductance $G$ of a one-dimensional channel to diverge as its length approaches zero. This would not be a surprise since the influence of scattering vanishes as the dimension reduces significantly.  However, even for a perfect conducting channel, $G$ has a finite value, $G_{Q}=2e^{2}/h$, known as the \textit{quantum of conductance} \cite{Kittel2004, Datta_1995}. 
Alternatively, this can be characterized in terms of its reciprocal $ R_{Q}\approx 12.906\hspace{0.05cm}\text{k}\Omega $.
If the channel is not a perfect conductor, which can be interpreted as having a single barrier, the probability for an electron with a Fermi energy $\epsilon_{f}$ to pass through it is $\mathcal{T}(\epsilon_{f})$. The conductance of the channel is then expressed by the \textit{Landauer formula} \cite{landauer5392683} 
\begin{equation}
    G(\epsilon_{f})=(2e^{2}/h)\mathcal{T}(\epsilon_{f}),
\label{land01}
\end{equation}
where for multiple channels, $\mathcal{T}(\epsilon_{f})=\sum_{n,m}\mathcal{T}_{n,m}(\epsilon_{f})$ with $n$ and $m$ denoting the transversal states (\ref{ring15}).

In the situation depicted in Fig. \ref{ringdevice}a, two leads are weakly coupled to the ring. They connect the device to two reservoirs and the contact points with the ring act as a double barrier (two barriers in a series). The system has a non-zero temperature of $40\hspace{0.05cm}\text{mK}$ and electrons are subject to loose phase coherence while traveling through the conductor path due, for instance, to scattering of electrons by phonons \cite{Kittel2004, Büttiker1991, FENG1990439, PhysRevB.44.13148}, impurities \cite{PhysRevB.18.5637}, electron-electron interactions \cite{doi:10.1142/S0217984902004032} and interfacial roughness \cite{Vinter1991, 10.1063/1.98305}.

Despite that, the decoherence resulting from inelastic scattering in this temperature regime is small enough that we can still expect to observe the effects of phase coherence. Therefore, the Landauer formula should be rewritten as
\begin{equation}
    G(\epsilon_{f})=-(2e^{2}/h)\sum_{n,m}\int \mathcal{T}_{n,m}(E)\dfrac{d}{dE}f(E-\epsilon_{f})dE,
    \label{land02}
\end{equation}
where $f(E-\epsilon_{f})$ is the Fermi-Dirac distribution and the integral is taken in an arbitrarily small interval centered at $\epsilon_{f}$ to avoid states that contribute no net current upon application of an electric field through the conductor \cite{Datta_1995}. When $T=0\hspace{0.05cm}\text{K}$, we recover Eq. (\ref{land01}). Our aim now is to obtain an expression for $\mathcal{T}_{n,m}(E)$ which will be, in general, a combination of both coherent and incoherent transmissions, $\mathcal{T}_{c}$ and $\mathcal{T}_{d}$ respectively.

In a double-barrier configuration, as the width of the well $W$ becomes smaller, the system acts as a filter, allowing states with a specific energy to pass through the barriers with very high probability even if their transmission coefficients combined are nearly zero. If each barrier has transmission probabilities $T_{1}$ and $T_{2}$, reflection probabilities $R_{1}\approx 1$ and $R_{2}\approx 1$, and if an inelastic scattering factor $\exp(-2\gamma W)$ has to be included every time the electron traverses the ring path, the coherent transmission coefficient in the regime of $\gamma \ll W^{-1}$ (low scattering) near the resonant condition is given by \cite{Datta_1995}
\begin{equation}
\mathcal{T}_{c}(E)=\dfrac{\Gamma_{1}\Gamma_{2}}{(E-E_{n,m})^{2}+(\Gamma_{1}+\Gamma_{2}+\Gamma_{\phi})^{2}/4},
\label{land03}
\end{equation}
where, if the electrons are reflected back and forth between the barriers with a velocity $v$,
\begin{equation}
\dfrac{\Gamma_{1}}{\hbar}=T_{1}\nu, \quad \dfrac{\Gamma_{2}}{\hbar}=T_{2}\nu\quad\text{and}\quad \dfrac{\Gamma_{\phi}}{\hbar}=(4\gamma W)\nu.  
\label{land04}
\end{equation} 
In this expression, $\nu=v/2W$ is the frequency of an electron impinging on one of the barriers and $2\gamma$ denotes the rate of inelastic scattering per round-trip, $2W$. The quantities $\Gamma_{1}$ and $\Gamma_{2}$ are known as the elastic broadenings, and $\Gamma_{\phi}$ is referred as the inelastic broadening. Thus, $\Gamma_{1}/\hbar$ and $\Gamma_{2}/\hbar$ can be interpreted as the number of times per second an electron successfully passes through the two barriers. Similarly, $\Gamma_{\phi}/\hbar$ is the number of times per second the particle is inelastically scattered along the path.
\begin{figure}[t]
\centering
\includegraphics[width=0.5\textwidth]{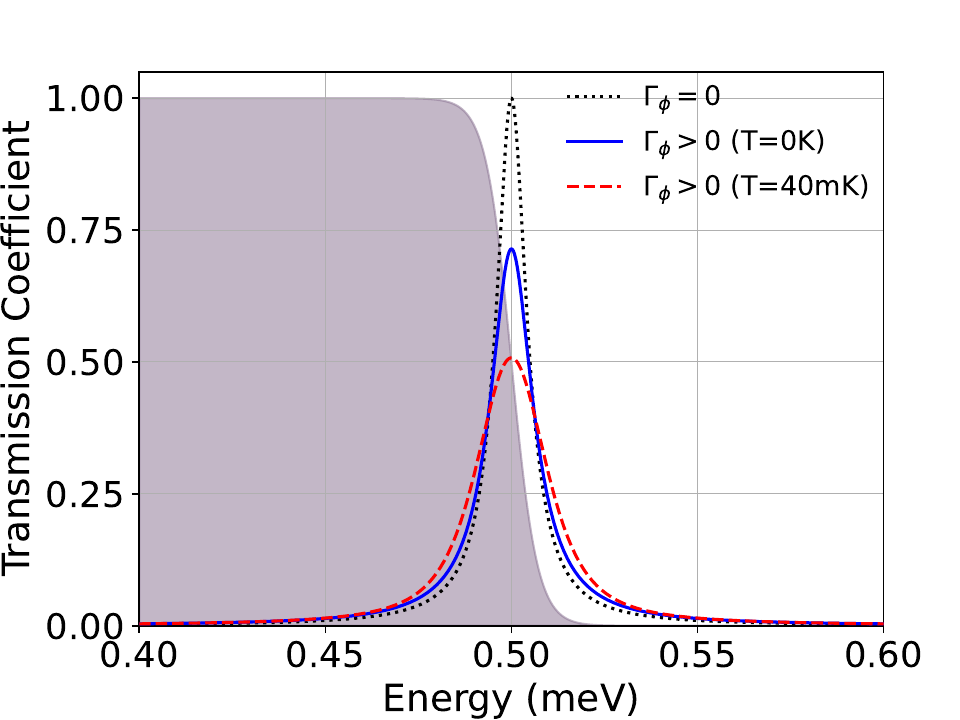}
\caption{(Color online) Plot of the expression for the transmission coefficient per energy state given by Eqs. (\ref{land02}) and (\ref{land06}). We have set $\Gamma_{1}=\Gamma_{2}=0.005\hspace{0.05cm}\text{meV} $ and $\Gamma_{\phi}=0.004\hspace{0.05cm}\text{meV}$. The shaded area represents the Fermi-Dirac distribution at $40 \hspace{0.05cm}\text{mK}$.}
\label{resonant}	
\end{figure}

The incoherent component of the transmission, on the other hand, is written as \cite{Büttiker1991}
\begin{equation}
    \mathcal{T}_{d}(E)=\dfrac{\Gamma_{\phi}}{(\Gamma_{1}+\Gamma_{2})}\Biggr\{\dfrac{\Gamma_{1}\Gamma_{2}}{(E-E_{n,m})^{2}+(\Gamma_{1}+\Gamma_{2}+\Gamma_{\phi})^{2}/4}\Biggr\}.
    \label{land05}
\end{equation}
Thus, we can compute the complete transmission coefficient for each channel, $\mathcal{T}_{n,m}(E)$, as
\begin{equation}
\mathcal{T}_{n,m}(E)=\dfrac{\Gamma_{1}\Gamma_{2}}{(\Gamma_{1}+\Gamma_{2})}\dfrac{(\Gamma_{1}+\Gamma_{2}+\Gamma_{\phi})}{[(E-E_{n,m})^{2}+(\Gamma_{1}+\Gamma_{2}+\Gamma_{\phi})^{2}/4]}.
\label{land06}
\end{equation}
Finally, Eqs. (\ref{land03}) and (\ref{land05}) imply that the correspondent parcels from the coherent and incoherent transmissions related to the total transmission are, respectively, $(\Gamma_{1}+\Gamma_{2})/\Gamma$ and $\Gamma_{\phi}/\Gamma$ with $\Gamma=\Gamma_{1}+\Gamma_{2}+\Gamma_{\phi}$. 

In Fig. \ref{resonant}, we show the complete transmission coefficients in terms of the energy for both coherent and sequential ($\Gamma_{\phi}>0$) tunneling. As expected, we obtain the shape of a Lorentzian with a peak centered at the Fermi level of $0.5\hspace{0.05cm}\text{meV}$. When there is only elastic scattering inside the conductor, electrons near $\epsilon_{f}$ have $\approx 100\%$ transmission probability. As we consider inelastic scattering and temperature effects in the system, the peak shifts downward and the Lorentzian broadens, increasing the probabilities for those states further from $\epsilon_{f}$. If we set $\Gamma_{1}=\Gamma_{2}=0.005\hspace{0.05cm}\text{meV} $ and $\Gamma_{\phi}=0.004\hspace{0.05cm}\text{meV}$ at $40 \hspace{0.05cm}\text{mK}$, the transmission peak decays to $\approx 50\%$. 

In the context of the conical geometry, the elastic broadenings must suffer a correction of $1/\alpha$ due to the influence of curvature in width $W$ between the barriers. However, no influence should be observed in the inelastic broadening since it has an increase in attempt frequency. 

\section{Results}\label{result}

To obtain and evaluate the results related to the electric charge transport in the system, we consider the parameter values used in Ref. \cite{PhysRevB.53.6947} to mimic the experimental device developed from Ref. \cite{PRB.1993.48.15148}. We then set $\hbar\omega_{0}=0.4496\hspace{0.05cm}\text{meV}$ and $r_{0}=800\hspace{0.05cm}\text{nm}$. Thus, we have a flat ring width $\Delta r=300\hspace{0.05cm}\text{nm}$ for a Fermi energy of $2 \hspace{0.05cm}\text{meV}$ and $\Delta r=150\hspace{0.05cm}\text{nm}$ for $\epsilon_{f}=0.5 \hspace{0.05cm}\text{meV}$. In a GaAs sample, the electron effective mass is $\mu=0.067 m_{e}$, where $m_{e}$ is the usual electron mass. Furthermore, in the context of an experimental temperature of $40 \hspace{0.05cm}\text{mK}$ and a weak field ($\omega_{c}<<\omega_{0}$), we set the elastic broadenings to $\Gamma_{1}=\Gamma_{2}=(0.005/\alpha)\hspace{0.05cm}\text{meV}$. This choice is in agreement with \cite{PhysRevB.53.6947} and can be justified since significantctrons have energy $\hbar\nu$ of tof order of $\approx 0.1\hspace{0.05cm}\text{meV}$, implying that $T_{1}$ and $T_{2}$ are both of tof order of $0.01$. 

We ignore, for simplicity, both dependencies on the magnetic field and quantum numbers for $\Gamma_{1}$, $\Gamma_{2}$ and $\Gamma_{\phi}$ \cite{PhysRevB.53.6947} as we can verify in Figs. \ref{subbands} and \ref{resonant} that the most significant states to the charge transport (near $\epsilon_{f}$) have low difference in energy when the magnetic field is changed. Note that if $T_{1}=T_{2}=0.01$, the probability of an electron being inelastically scattered through a round trip would be approximately $4\gamma W$=0.008, which seems to be a high chance compared to the barriers transmission coefficients combining for $0.0001$. However, in the resonant condition (\ref{land06}), we are concerned with states near $\epsilon_{f}$ which are able to carry net current so that for the significant contributions in the transmission, $\mathcal{T}_{n,m}>4\gamma W$. 

\subsection{Conductance and Van-Hove Singularities}

\begin{figure}[!h]
\centering
\begin{subfigure}{0.5\textwidth}
\includegraphics[width=\textwidth]{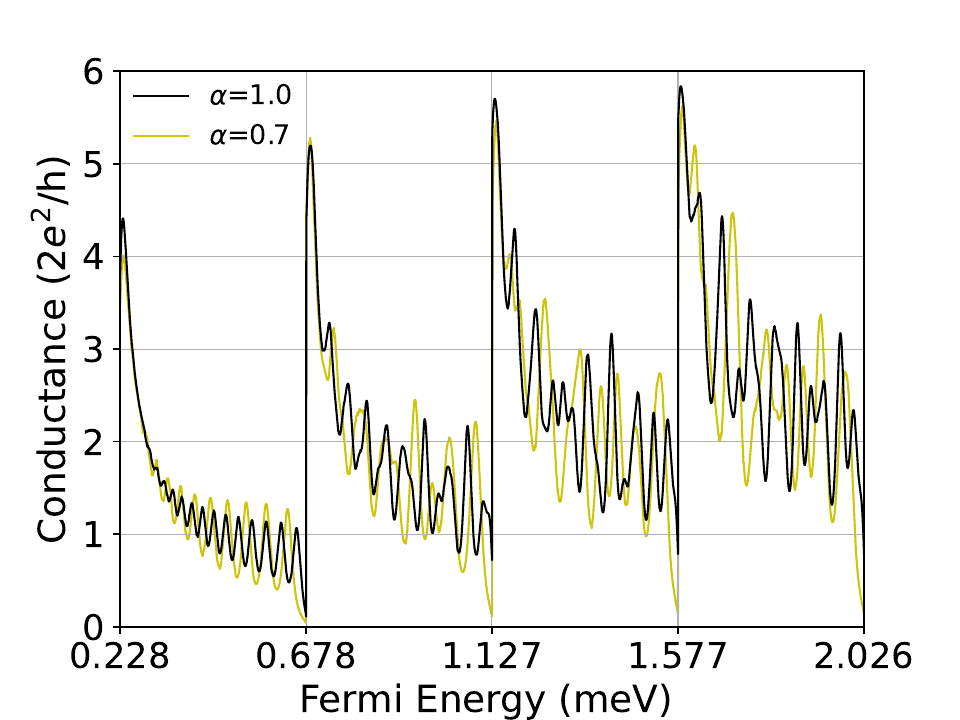}
\subcaption{}
\label{vanhoove}
\end{subfigure}
\begin{subfigure}{0.5\textwidth}
\includegraphics[width=\textwidth]{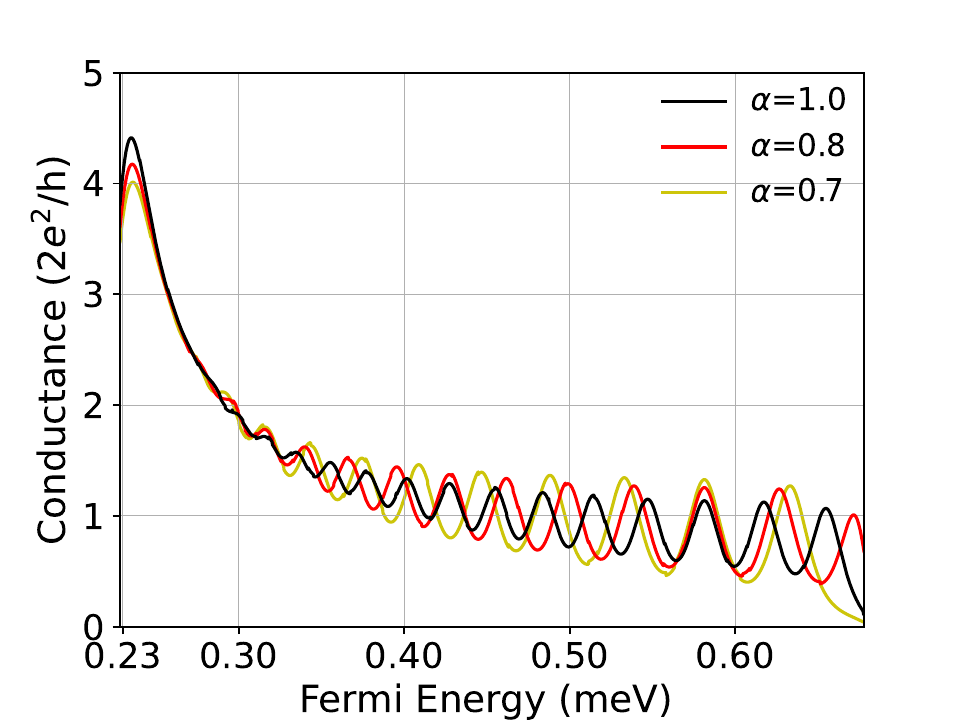}
\subcaption{}
\label{fermiconduc}
\end{subfigure}
\caption{(Color online) (a) Conductance as a function of the Fermi energy for flat and curvature regime. The peaks result from the density of states per unity of energy \cite{Kittel2004}. (b) Enlarged plot of Fig. \ref{vanhoove} for the first subband.}
\label{vanhoovesing}
\end{figure}

Fig. \ref{vanhoove} shows the conductance behavior as a function of the Fermi energy for the cases of a curved and a flat ring. This plot is deeply relevant for experimental characterization on the device since $\epsilon_{f}$ can be set by a gate voltage that controls the number of electrons occupying the system \cite{PRB.1993.48.15148, 10.1063/1.881503, PhysRevLett.60.848}. The peaks result from the high density of states near the bottom of each subband (see Fig. \ref{subbands}) and are known as the Van-Hove singularities \cite{Kittel2004}. The total conductance curve then performs an oscillating decay until the Fermi level reaches the bottom of the following subband, generating another peak. Thus, one can directly measure, to high accuracy, the quantity $h\omega_{0}$ that characterizes the confinement. The noisy oscillations after the second singularity result from the complex transitions due to the superposition of states from different $n$ levels.

As we have seen in Figs. \ref{subband1} and \ref{subband07}, the conical geometry reduces the density of states per unity of energy and increases the gap between state crossing points. Thus, from Eq. (\ref{land06}), both amplitudes and periods of the conductance oscillations should grow with a reduction in $\alpha$ as is clearly noticeable in Fig. \ref{fermiconduc}. The Van-Hove singularities are also affected by a downward shift of their peaks as the curvature parameter decreases. Note that the average value of the conductance always increases from one peak to another due to the cumulative contributions from the lower subbands.

\subsection{Magnetoresistance}

\begin{figure*}[t!]
\centering
\begin{subfigure}{0.4\textwidth}
\includegraphics[width=\textwidth]{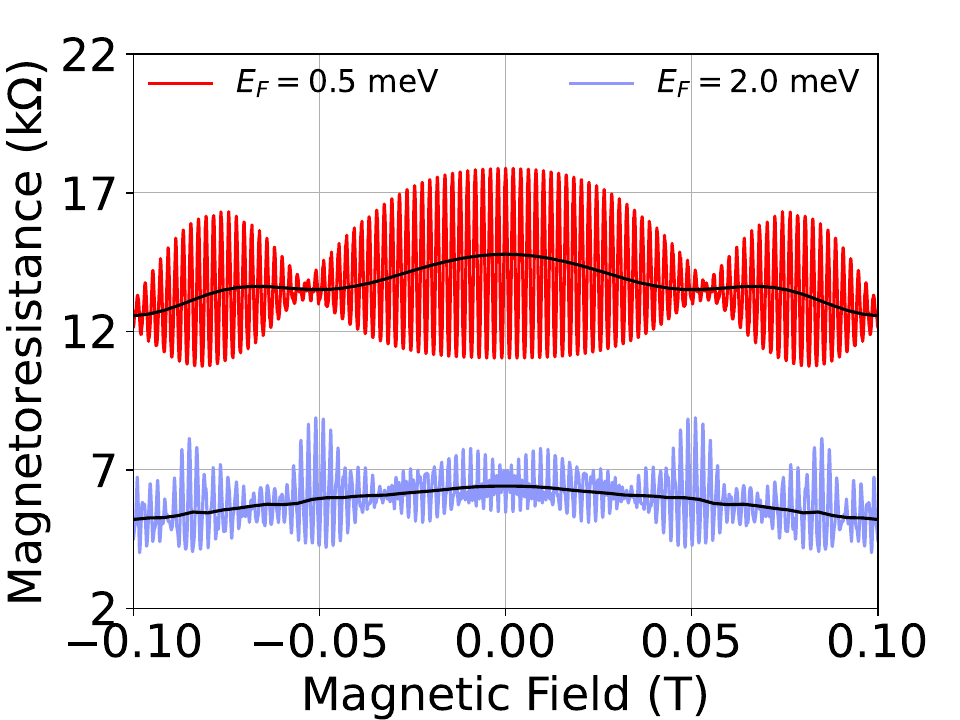}
	\subcaption{$ \alpha=1.0 $}
	\label{magnetoresistance1}
\end{subfigure}
\begin{subfigure}{0.4\textwidth}
	\includegraphics[width=\textwidth]{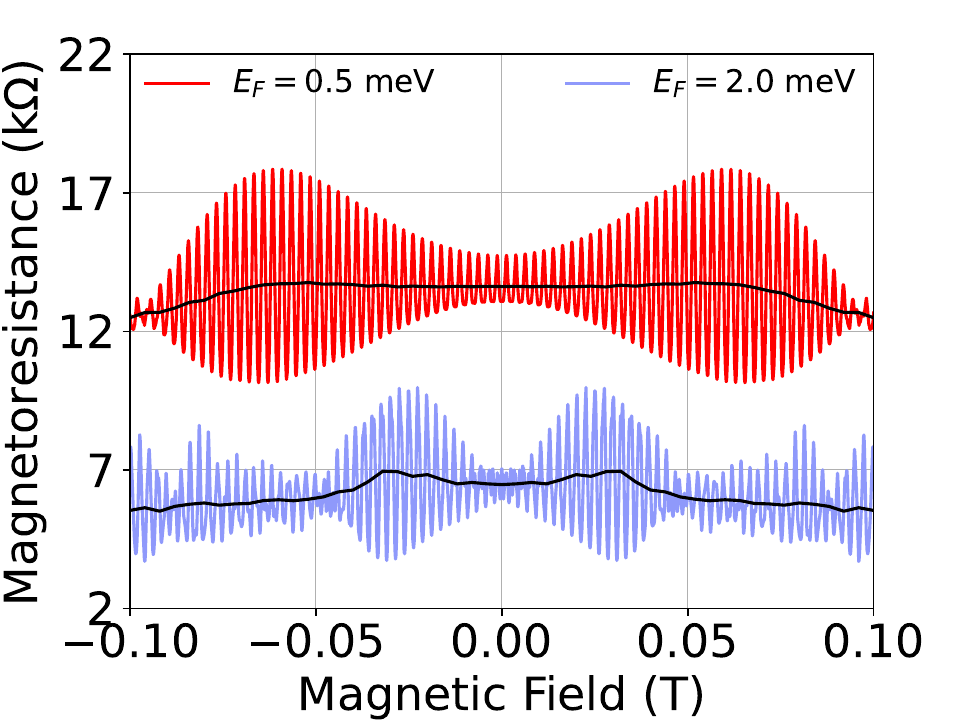}
	\subcaption{$ \alpha=0.9 $}
    \label{magnetoresistance09}
\end{subfigure}
\begin{subfigure}{0.4\textwidth}
	\includegraphics[width=\textwidth]{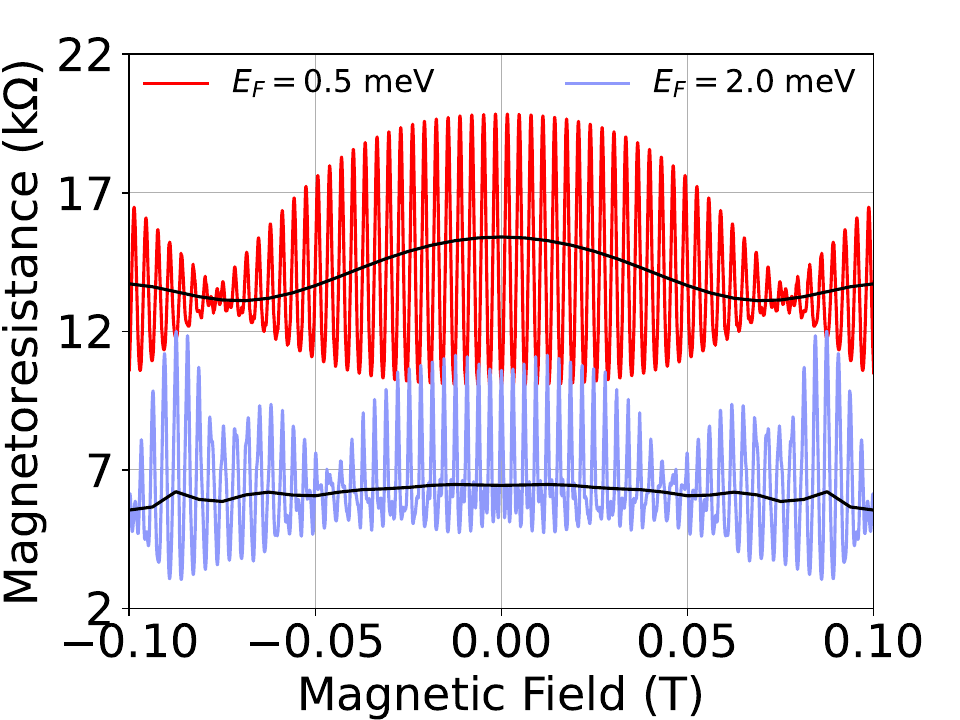}
	\subcaption{$ \alpha=0.8 $}
	\label{magnetoresistance08}
\end{subfigure}
\begin{subfigure}{0.4\textwidth}
	\includegraphics[width=\textwidth]{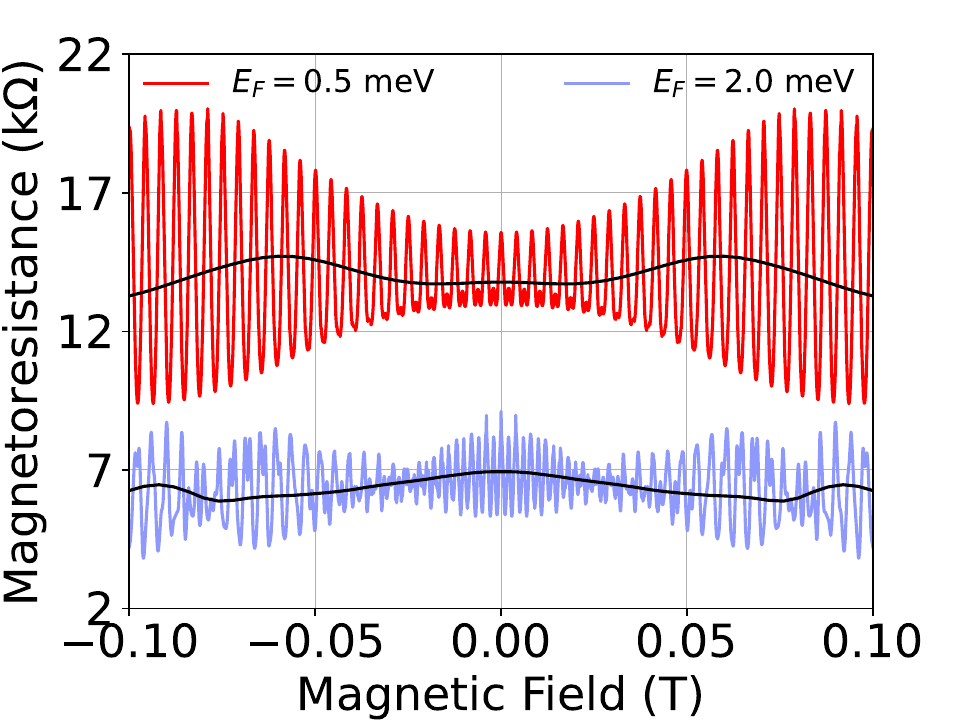}
	\subcaption{$ \alpha=0.7 $}
	\label{magnetoresistance07}
\end{subfigure}
\caption{(Color online) Magnetoresistance as a function of the magnetic field for different values of curvature parameter $\alpha$. The average curve points were obtained using intervals of $5\hspace{0.05cm}\text{mT}$.}
\label{resistancemag}
\end{figure*}

We now turn to analyze the magnetoresistance of the system. In Fig. \ref{resistancemag}, we plot its behavior with the magnetic field for different values of curvature parameter and two regimes of subband occupation: one ($0.5 \hspace{0.05cm}\text{meV}$) and four ($2 \hspace{0.05cm}\text{meV}$). One can identify an interference pattern distinctive from a beating effect for the lower energy case (red curve). In Figs. \ref{subband1} and \ref{subband07}, we can note that the energy of the states can shift upward or downward with $B$ as a result of the clockwise or anti-clockwise directions of motion of the electrons as they resist the change in the flux of magnetic field through the internal surface defined by the ring \cite{PhysRevB.53.6947, PhysRevB.76.035110}.

As discussed previously, near $\epsilon_{f}$, the contribution from each state to the magnetoresistance is an AB oscillation with period $\approx 2.037\hspace{0.05cm}\text{mT}$. However, their periods acquire a slight difference with the variation in $n$ and $B$ \cite{PhysRevB.53.6947}. Therefore, since we have only two sets of oscillations for one occupied subband, the sum of all the contributions results in the characteristic beating pattern presented. Since we have four occupied levels for the higher energy case (light blue curve), the superposition of eight sets of oscillations providing state transitions between different subbands will result in apparently noisy oscillations and complex envelope patterns \cite{PhysRevB.53.6947}.

The effects of curvature in this property are noticeable by following the behavior of the envelope functions. The conical geometry increases the maximum amplitudes of the AB oscillations (antinodes) and their period by $\approx p_{0}/\alpha^{2}$. These amplitude peaks occur when the Fermi energy matches, in a certain interval of the magnetic field, the curve connecting state crossing points with very similar energies (Fig. \ref{subband1}). On the other hand, the lowest amplitude regions represented by the nodes occur when $\epsilon_{f}$ is located in the very middle region between these curves (Fig. \ref{subband07}) and are dominated by oscillations with period $p_{0}/2\alpha^{2}$ \cite{PhysRevB.76.035110}.

As we see by comparing Figs. \ref{subband1} and \ref{magnetoresistance1}, the $0.5\hspace{0.05cm}\text{meV}$ dashed line nearly passes through the curve defined by the state crossing points when $0\hspace{0.05cm}\text{T}$ and, thus, we have an antinode. A similar situation happens to $\alpha=0.8$ in Fig. \ref{magnetoresistance08}.  Conversely, comparing Figs. \ref{subband07} and \ref{magnetoresistance07} for the case of $\alpha=0.7$, Fermi energy is located approximately between these curves, providing an amplitude node, for which a similar pattern can be observed in Fig. \ref{magnetoresistance09}. These pattern recurrences can be interpreted as the first indication of a periodic dependence in the magnetoresistance with $\alpha$.

More evidence of this dependence can be found by analyzing the averaged magnetoresistance (solid black curve) in Fig. \ref{resistancemag} obtained through intervals of $5\hspace{0.05cm}\text{mT}$, where we can identify an oscillatory
variation of this quantity with $\alpha$ for several values of $B$. Furthermore, it is important to note that the most stable regimes for the averaged magnetoresistance are carried by the nodes, whereas the local peaks in the curve localize the antinodes with good precision.

Looking directly at the $\epsilon_{f}= 2\hspace{0.05cm}\text{meV}$ regime, we highlight
that, as Figs. \ref{subbandsmin} and \ref{subbandsmin07} suggest Fermi energy is near the bottom of the next subband and, even though we are not concerned with the current itself, if one decides to apply a weak and constant electric field along the ring path, the variation in the momentum of an electron, $\Delta\braket{\hat{p}}\approx\Delta m(\hbar/r_{n,m})$ happens only through the variation in angular quantum number \cite{Datta_1995}. Note that we have not considered the variation in  $r_{n,m}$ since $L$ has no significant change for small displacements $\Delta m$ occurring in the same subband. Thus, the electrons can not access a different level through just an applied voltage, and the states from the subband $n=5$ are not allowed to be occupied, as well as the $n\le4$ ones with energy above its bottom since they can only be reached by transitions through the states in this forbidden subband. For that reason,  we should not take these states into account in Eq. (\ref{land06}) and, therefore, the effect of curvature in the bottom of this subband plays an important role in the behavior of the magnetoresistance shown in Fig \ref{resistancemag}.

\subsection{Magnetoresistance Oscillations in $\alpha$}

\begin{figure}[t]
\centering
\begin{subfigure}{0.5\textwidth}
    \includegraphics[width=\textwidth]{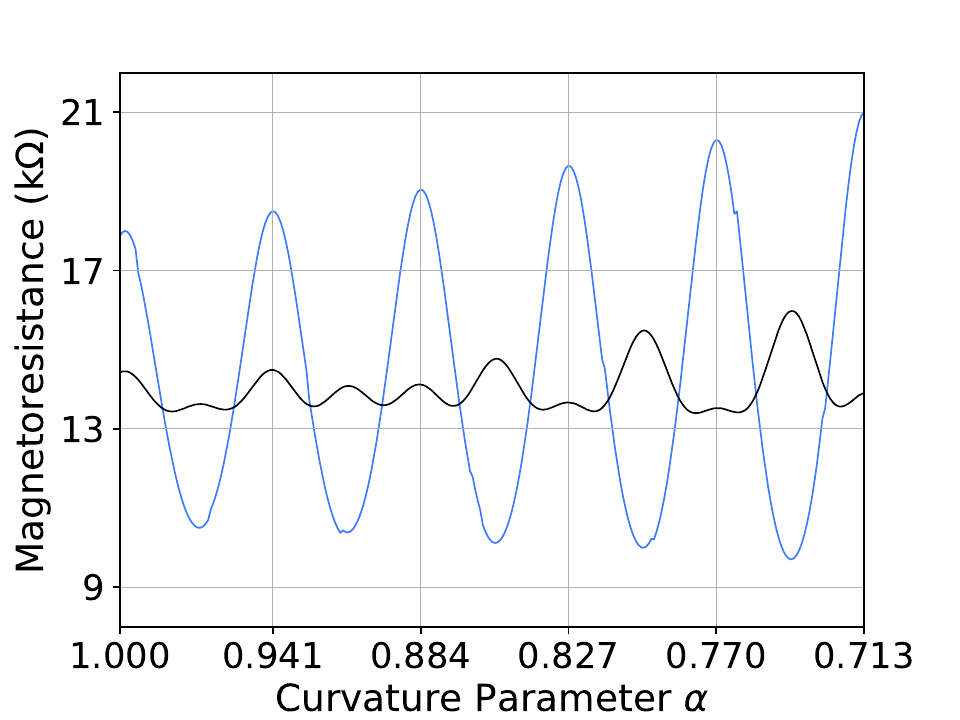}
    \subcaption{}
    \label{oscillationresistance}
\end{subfigure}
\begin{subfigure}{0.5\textwidth}
    \includegraphics[width=\textwidth]{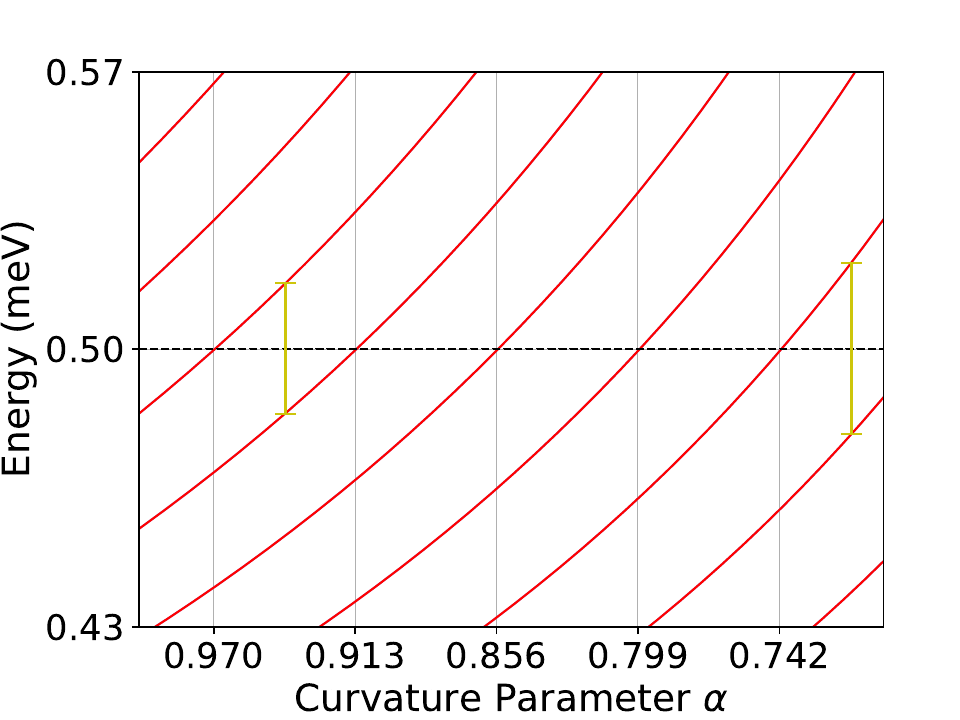}
    \subcaption{}
    \label{statecurvature0}
\end{subfigure}
\caption{(Color online) (a) Magnetoresistance of the system at $0\hspace{0.05cm}\text{T}$ (light blue) and its average (dark) as a function of the curvature parameter. The characteristic period of oscillation in $\alpha$ is approximately $0.057$. (b) Energy eigenvalues as a function of the curvature parameter. The crossing points between the energy curves and Fermi level localize the conductance peaks in (a). The yellow vertical bars represent the energy gap between the energetic neighbor states nearest $\epsilon_{f}$ in the configuration of the local maximum for the resistance.}
\label{curvatureoscillation0}
\end{figure}

In Fig. \ref{oscillationresistance}, we present an almost periodic behavior of the magnetoresistance with the curvature for $B=0$ (light blue curve). The period of these oscillations $\Delta\alpha$ is approximately $0.057$, which refers to a $\approx 20.52^{\circ}$ angular deficit in the material \cite{DELIRA2024115898}. After a slight increase in the resistance until $\alpha\approx0.998\hspace{0.05cm} (0.72^{\circ})$, the conductance grows until it achieves its first peak at $\alpha\approx 0.970$ $(10.8^{\circ}$). Thus, the charge transport through the ring device can be optimized through specific regimes of deformations in the material structure. Each peak in the oscillating pattern is due to the Fermi energy being energetically in the middle between two state crossing points. On the other hand, $\epsilon_{f}$ is located exactly in a state crossing point for the valleys, at which we have the configuration where conductance is maximum.

\begin{figure}[t!]
\centering
\begin{subfigure}{0.5\textwidth}
\includegraphics[width=\textwidth]{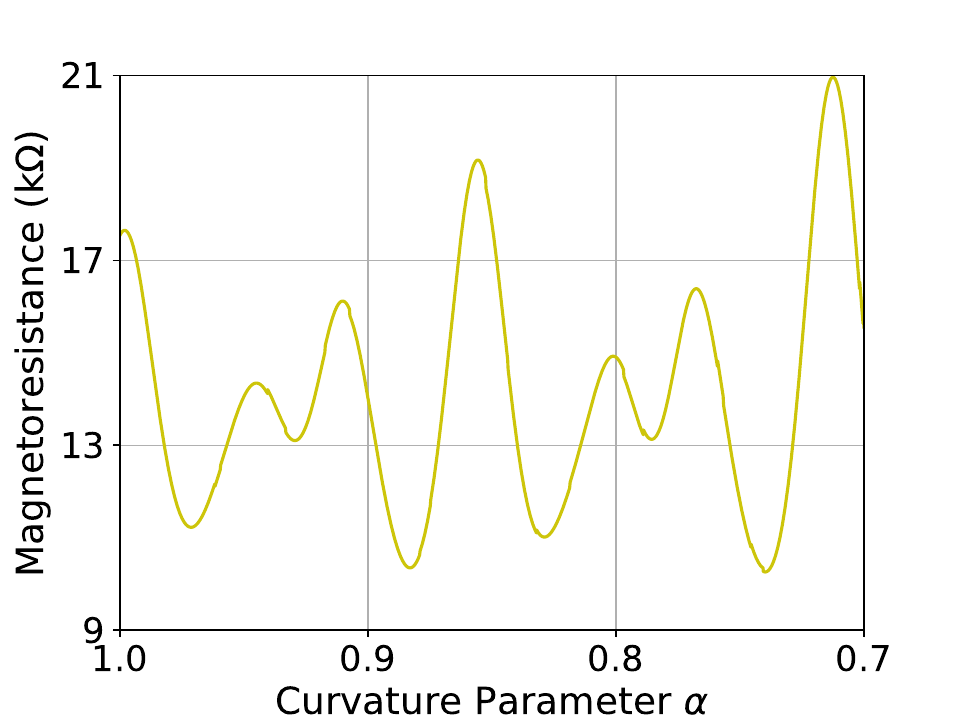}
\subcaption{}
\label{oscillationresistance2}
\end{subfigure}
\begin{subfigure}{0.5\textwidth}
\includegraphics[width=\textwidth]{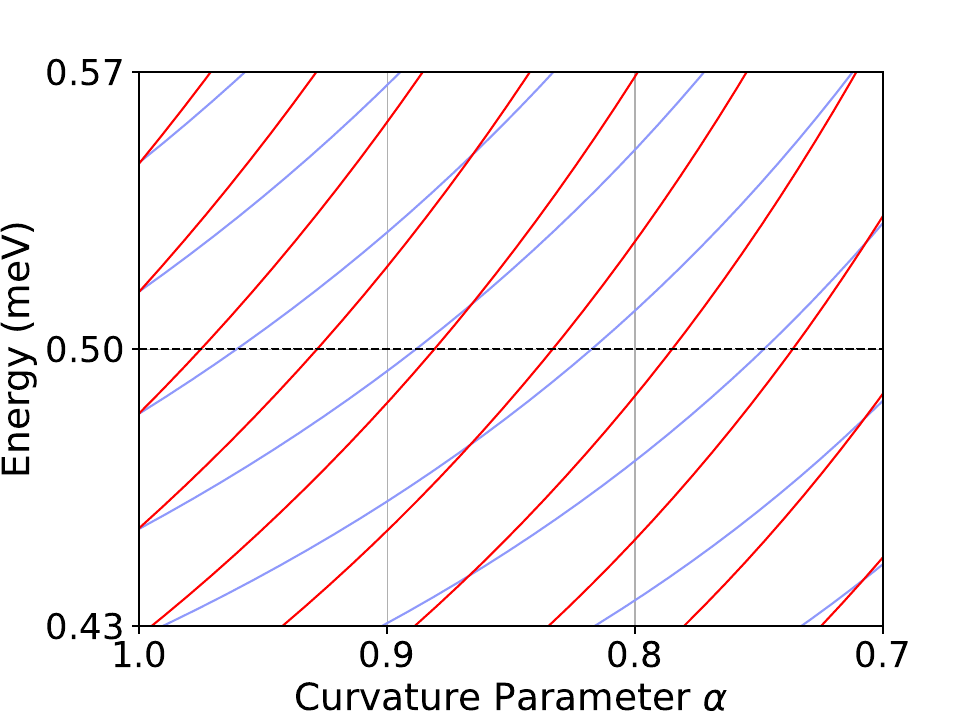}
\subcaption{}
\label{statecurvature2}
\end{subfigure}
\caption{(Color online) (a) Magnetoresistance as a function of the curvature parameter for $2p_{0}\approx 4.074\hspace{0.05cm}\text{mT}$. (b) Energy eigenvalues as a function of the curvature parameter for $2p_{0}$. The red curves represent the $m\ge 0$ states, and the light blue curves the $m<0$ states.} 
\label{curvatureoscillation2}
\end{figure}

\begin{table*}[t!]\centering
\setlength{\tabcolsep}{12 pt}
\vspace{0.15 cm}
\caption[12pt]{Curvature parameter values for significant configurations of the magnetoresistance oscillations at $B=0$ }\label{tab1}
\begin{tabular}{cccccc}
\hline
\hline
\multicolumn{1}{c}{}
& \multicolumn{1}{c}{$\alpha_{1}$} & \multicolumn{1}{c}{$\alpha_{2}$}  & \multicolumn{1}{c}{$\alpha_{3}$} & \multicolumn{1}{c}{$\alpha_{4}$} & \multicolumn{1}{c}{$\alpha_{5}$} \\
\hline
Resistance Peaks	& $ 0.998 $ & $0.941$ & $0.884$ & $0.827$ & $0.770$\\
Conductance Peaks	& $ 0.970 $ & $0.913$ & $0.855$ & $0.798$ & $0.741$\\
Amplitude Anti-nodes & $ 0.998 $ & $0.970$  & $0.941$ & $0.913$ & $0.884$ \\
Amplitude Nodes & $ 0.984 $ & $0.956$ & $0.927$ & $0.899$ & $0.870$ \\
\hline
\hline
\end{tabular}	

\end{table*}

The periodicity can be explained by the intersections of the Fermi level with the energy eigenvalues as we vary $\alpha$. This can be visualized in Fig. \ref{statecurvature0}, where we also can identify an enlarging rate for gaps between energetic neighbor states around $\epsilon_{f}$. This is responsible for increasing the amplitudes of the resistance (i.e., the lift in the peaks). The amplitudes of conductance, however, also increase due to the effect of curvature in the elastic broadenings.

The averaged magnetoresistance (black curve) also provides us with an oscillating pattern for the antinodes (peaks), which, compared with Fig. \ref{resistancemag}, is in good agreement with our previous assumption that the antinodes carry the local maximums of this quantity. The period of recurrence for antinodes is $\approx\Delta\alpha/2$. This periodicity is also valid for the nodes, which cannot be interpreted as the valleys in the same curve. The first node can be achieved for $\alpha\approx 0.984$ or $5.76^{\circ}$ in angular deficit, providing a reduction in the amplitude around $B=0$ of $92.6\%$ compared to the flat case. Thus, configurations of stable resistance against external magnetic field fluctuations can be achieved by tuning the curvature parameter.

In the situation where the magnetic field is not zero, Fig. \ref{statecurvature2} show us that, for $B=2p_{0}$, the energy eigenvalue curves are split into components of $m\ge0$ (red) and $m<0$ (light blue). The dynamics of intersection and interference between $\epsilon_{f}$ and these curves becomes complex, resulting in different patterns of oscillations presented in Fig. \ref{oscillationresistance2}, which are not periodic anymore. The increasing amplitudes are explained for the same reasons as the $B=0$ case.

\begin{figure}[t]
\centering
\begin{subfigure}{0.5\textwidth}
\includegraphics[width=\textwidth]{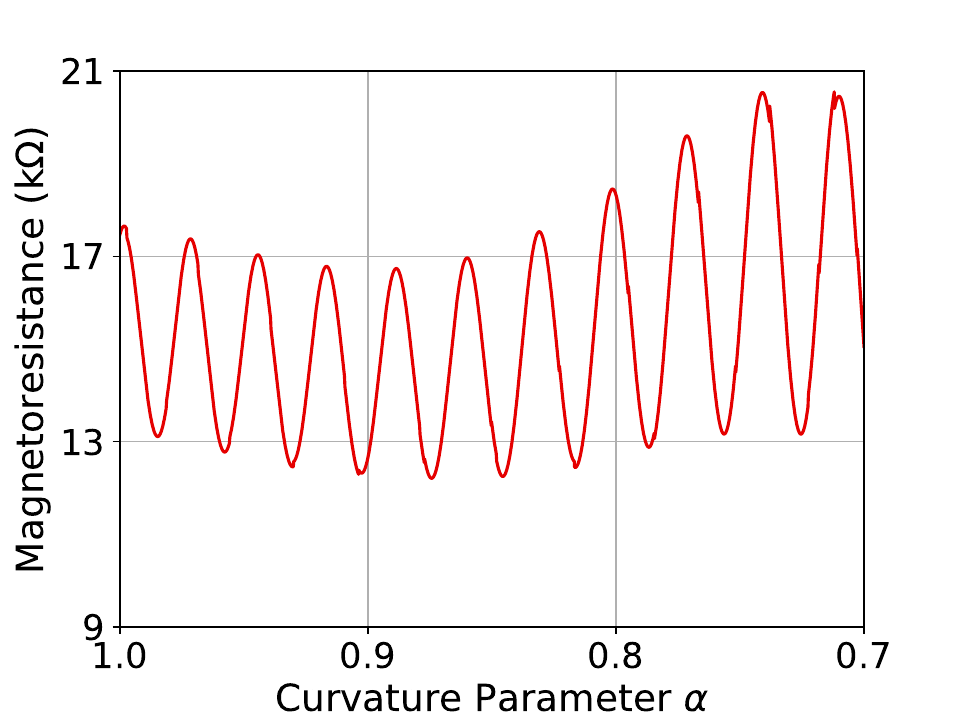}
\subcaption{}
\label{oscillationresistance10}
\end{subfigure}
\begin{subfigure}{0.5\textwidth}
\includegraphics[width=\textwidth]{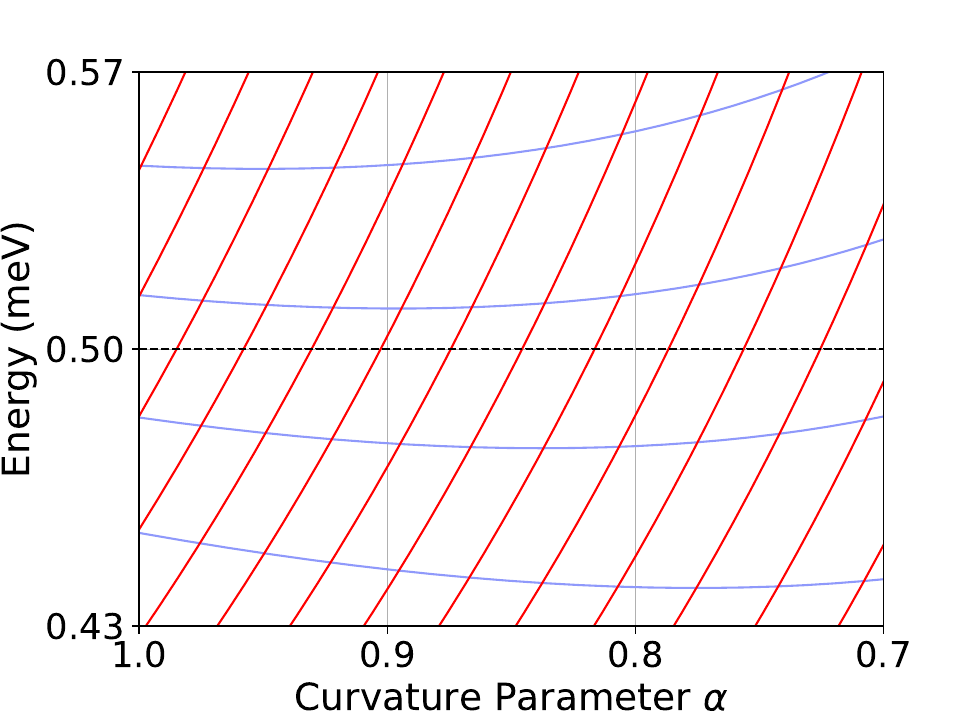}
\subcaption{}
\label{statecurvature10}
\end{subfigure}
\caption{(Color online) (a) Magnetoresistance as a function of the curvature parameter for $B=10p_{0}\approx 20.37\hspace{0.05cm}\text{mT}$. (b) Energy eigenvalues as a function of the curvature parameter for $B=10p_{0}$. The red curves represent the $m\ge 0$ states, and the light blue curves the $m<0$ states.} 
\label{curvatureoscillation10}
\end{figure}

As the value of $B$ intensifies to $10p_{0}$, the angle between $m\ge 0$ and $m<0$ curves (Fig. \ref{statecurvature10}) increases in such a way that only the first ones are crossed by the Fermi level until $\alpha=0.7$. Thus, oscillations are mainly governed by these intersections while the characteristic up and down behavior shown in Fig. \ref{oscillationresistance10} is determined by
both the approximation and the deviation of the $m<0$ energies from $\epsilon_{f}$. Furthermore, comparing Figs. \ref{statecurvature0}, \ref{statecurvature2} and \ref{statecurvature10}, we can identify that the number of intersections between the $0.5\hspace{0.05cm}\text{meV}$ line with the red curves determines the number of conductance peaks. Thus, the higher the value of $B$, the higher the number of intersections in a given interval of $\alpha$, and, consequently, more oscillations will be observed.

\subsection{Charge Current}

Since we have already discussed the resistance of the quantum ring, it is time to explore the charge transport through the application of an electric potential $V$ between its lower (L) and upper (U) terminals. In other words, we are now considering it as part of an electric circuit instead of an isolated system. This picture has led to a series of studies concerning the applications of mesoscopic devices as a nanoelectronic component \cite{MAITI2007199, He2020Analysis, Prasad2023Enhancement}.

To obtain the net current $I(V,T)$ for finite temperatures $T$, we have to take into account the difference between the Fermi-Dirac distributions between the terminals \cite{Kittel2004}. Thus, we have

\begin{equation}
    I(V,T)=\dfrac{2e}{h}\int^{\infty}_{-\infty} \{f_{L}(\epsilon-eV)-f_{U}(\epsilon)\}\mathcal{T}(\epsilon)d\epsilon.
\label{charge01}
\end{equation}
In the classical case $\mathcal{T}$ does not depend significantly on the energy of each electron state in such a way that 
\begin{equation}
    I(V)=\dfrac{2e}{h}\cdot\mathcal{T}\cdot (eV-0)= G \cdot V
\label{charge02}
\end{equation}
which gives us the well known Ohm`s Law. In order to explore the physics of Eq.(\ref{charge01}) in more detail, we separate our investigation into the following regimes

\begin{center}
    \textbf{Case $eV << \epsilon_{f}$}:
\end{center}

\begin{figure}[b!]
\centering
\includegraphics[width=0.45\textwidth]{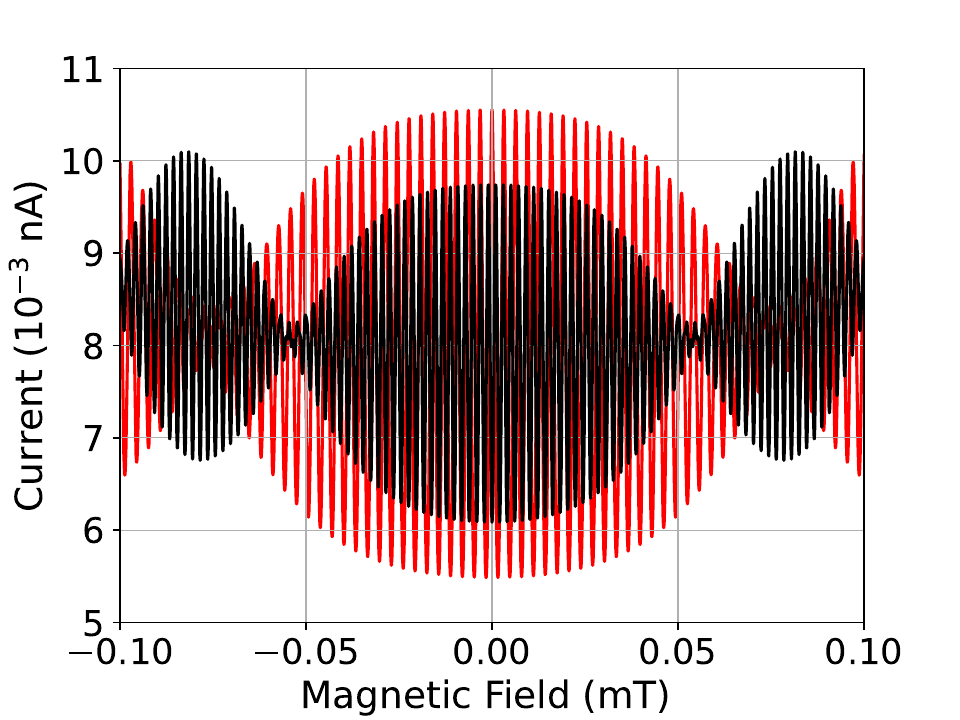}
\caption{(Color online) Plot of the result (\ref{charge02}) for extremely low voltages in terms of the external magnetic field for $\alpha=1$ (dark) and $\alpha=0.8$ (red).}
\label{currentmaglowgate}	
\end{figure}

For our purposes $\mathcal{T}=\mathcal{T}(\epsilon)$ and (\ref{charge02}) is just a tremendous simplification. Consider now the definition of the derivative of $f$ with respect to $\epsilon$ for small variations in the energy where $eV\leq k_{B}T<<\epsilon_{f}$ We have
\begin{equation}
    \lim_{-eV \rightarrow 0}\dfrac{f(\epsilon-eV)-f(\epsilon)}{(-eV)}= \dfrac{df}{d\epsilon}.
\end{equation}
Thus,
\begin{equation}
    \dfrac{f(\epsilon-eV)-f(\epsilon)}{eV} \approx -\dfrac{df}{d\epsilon}, \qquad (eV<<\epsilon_{f}).
    \label{charge04}
\end{equation}
Using (\ref{charge04}) in (\ref{charge01}),
\begin{equation}
    I(V,T)=-\dfrac{2e^2}{h}V \int^{\infty}_{-\infty} \dfrac{df}{d\epsilon}\mathcal{T}(\epsilon)d\epsilon = G \cdot V.
\label{charge05}
\end{equation}
This result shows us that, for small values of the voltage ($<<200\hspace{0.05cm}\mu\text{V}$) in the one-subband case, we still preserve Ohm's law at a quantum level!

\begin{figure}[t!]
\centering
\includegraphics[width=0.45\textwidth]{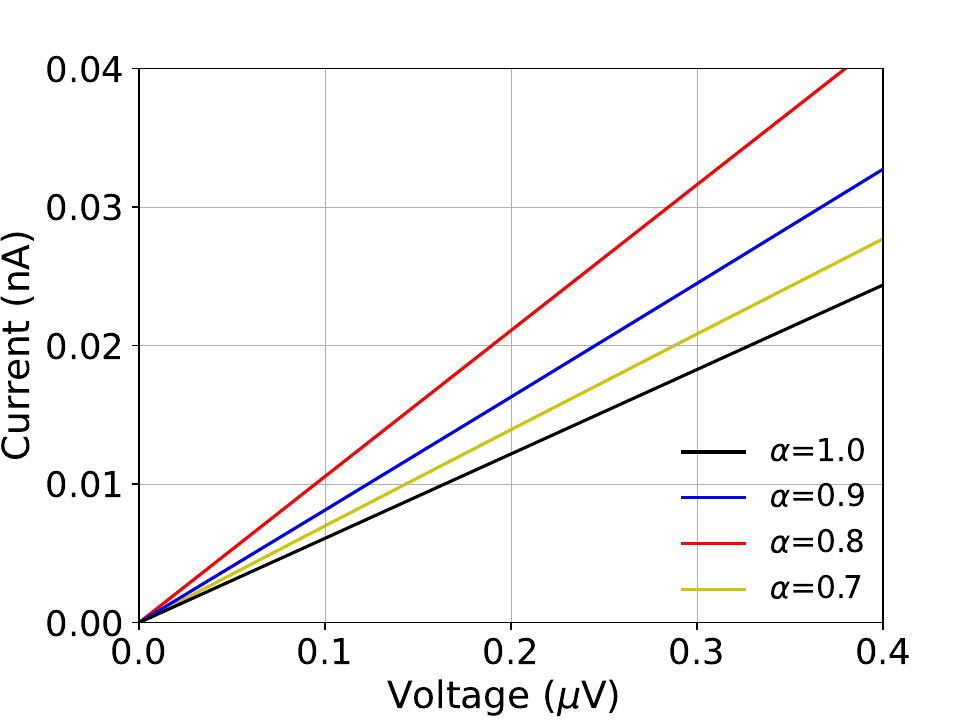}
\caption{(Color online) Plot of the result (\ref{charge02}) in terms of the extremely low voltage between the terminals for several values of the curvature parameter $\alpha$. The linear behavior evidences the characteristic Ohm's Law dependence.} 
\label{lowcurrent}	
\end{figure}

Figs. \ref{currentmaglowgate} and \ref{lowcurrent} show us numerically the consistence of this result. Fig. \ref{currentmaglowgate} can be seen simply as an inverse plot of the magnetoresistance presented in Figs \ref{magnetoresistance1} and \ref{magnetoresistance08} modulated by the extremely low voltage of $0.1\hspace{0.05cm}\mu\text{V}$, where the curvature contributes to amplify their amplitudes as previously discussed. 

Fig. \ref{lowcurrent} shows the behavior of the current with respect to voltage for several $\alpha$ cases without an external field. As predicted by Ohm's law, they have to perform a linear function where their inclination represents the inverse of the resistance. In other words, the lines with the lower angular coefficients have higher values for the resistance. This conclusion is in agreement with Fig. \ref{oscillationresistance}.

\begin{center}
    \textbf{Case $eV >> k_{B}T$}
\end{center}

For high voltage between the terminals ($eV >> K_{B}T= 0,344\hspace{0.05cm}\mu e\text{V}$), we can rewrite Eq. (\ref{charge01}) as 
\begin{align}
    I(V,T)=&\dfrac{2e}{h}\int^{\infty}_{-\infty} \Biggr\{\dfrac{1}{1+e^{\{\epsilon-(\epsilon_{f}+eV)\}/ k_{B}T}}\\
    & -\dfrac{1}{1+e^{(\epsilon-\epsilon_{f})/k_{B}T}}\Biggr\}\mathcal{T}(\epsilon)d\epsilon.
\label{charge06}
\end{align}
Since $\mathcal{T}(\epsilon)$ requires the states for the resonant tunneling to satisfy $\abs{\epsilon-\epsilon_{f}}< eV$, the first factor reduces to $1$ when $eV >> k_{B}T$. We then have
\begin{equation}  
    I(V,T)=\dfrac{2e}{h}\sum_{n,m}\int^{\infty}_{-\infty} \dfrac{\mathcal{T}_{n,m}(\epsilon)}{1+e^{-(\epsilon-\epsilon_{f})/k_{B}T}}d\epsilon.
\label{charge07}
\end{equation}

\begin{figure}[t!]
\centering
\includegraphics[width=0.48\textwidth]{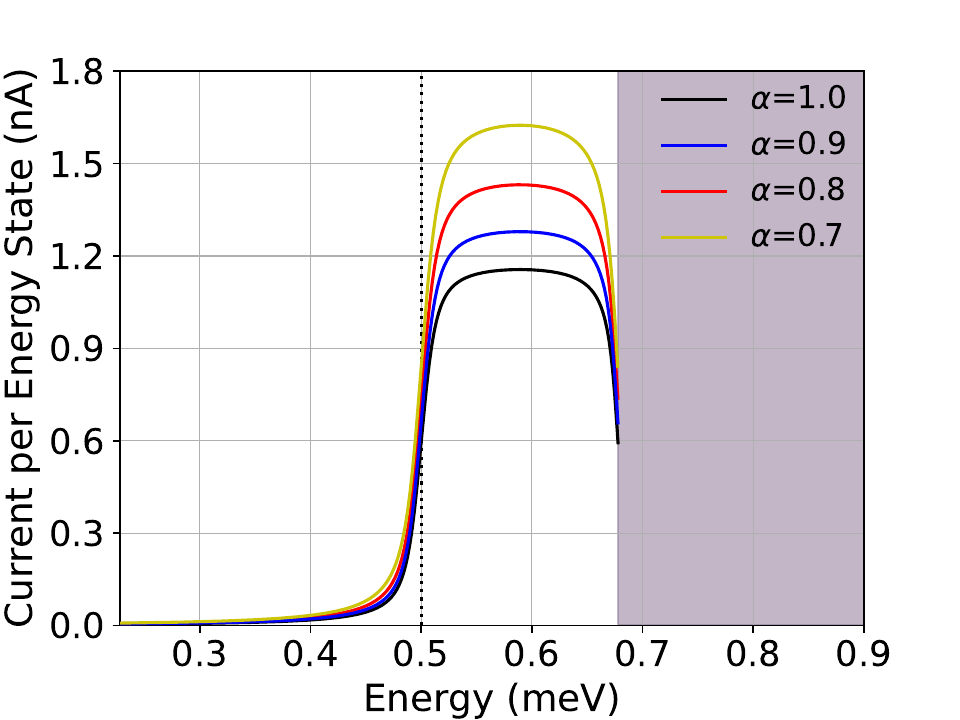}
\caption{(Color online) Plot of $I_{n,m}$ in the expression (\ref{charge07}) for the high-voltage regime. We see that states with energy higher than $\epsilon_{f}$ are the most contributors for the total net current. The shaded area represents the forbidden energetic region associated with the next subband.}
\label{currenttransmission}	
\end{figure}

In Fig. \ref{currenttransmission}, we present the behavior of the result (\ref{charge07}) for specific energy states $n$ and $m$. In other words, we calculate the current carried per energy state and the sum of them must give the total net current. An interesting point about this plot is that the most contributor states are localized above $\epsilon_{f}$ (dotted line), which is in agreement with the fact that electrons in energies higher than Fermi level are accomplished with great mobility! The shaded area represents the forbidden region of the next subband, where electrons cannot cross.

\begin{figure}[h]
\centering
\includegraphics[width=0.48\textwidth]{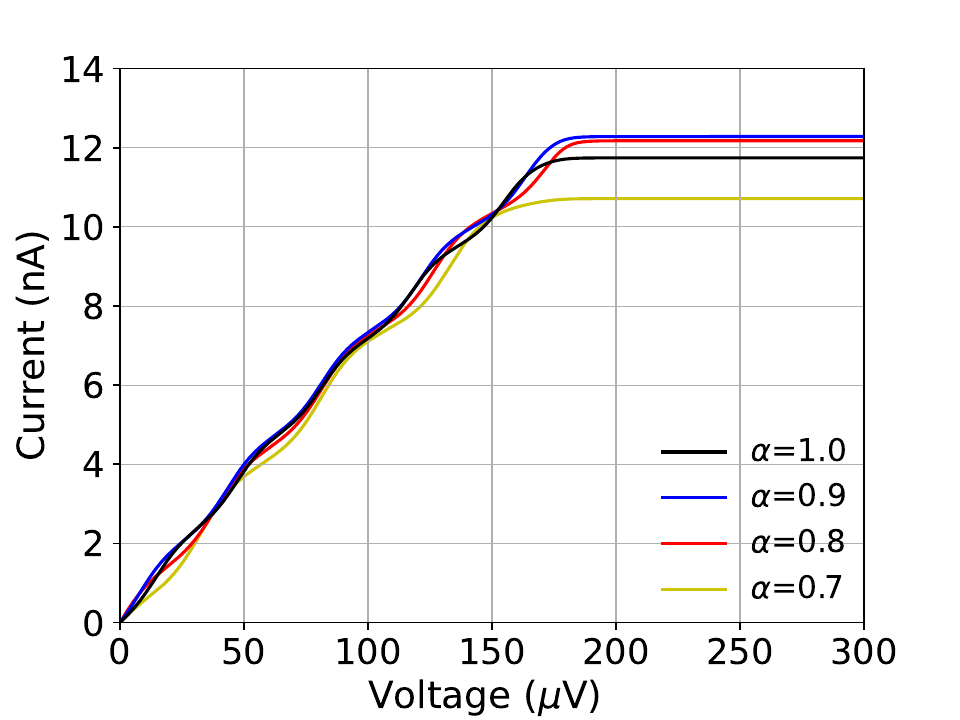}
\caption{(Color online) Plot of the general definition (\ref{charge01}) for $I(V,T)$. Under high-voltage, the system reaches a saturation point where the charge current presents a limit value regardless of the electric potential intensity.}
\label{currentV}	
\end{figure}

At this point, $I(V,T)$ reaches a saturation point where the voltage does not change the charge transport anymore. Thus, the current curve must exhibit a plateau profile for this limit case. This can be clearly observed in Fig. \ref{currentV} for voltages higher than $200\hspace{0.05cm}\mu\text{V}$. Therefore, the quantum ring could act as a device that controls the current in the circuit for strong oscillations of the electric field. We can note that the plateaus reach different intensities depending on the curvature parameter. The maximum value is presented for the case $\alpha=0.9$ and the minimum is presented for the case $\alpha=0.7$.

The respective limit values for the current in the different curvature regimes are not exactly followed by the results shown in Fig. \ref{currenttransmission}. The reason why the $\alpha=0.7$ system does not provide the highest value of the total net current is because the current per energy state is not the only factor we must take into account. As shown in Fig. \ref{subbands}, systems with a higher curvature have a lower density of states per unit of energy. Hence, for instance, the lower values of $I_{n,m}$ for $\alpha=0.9$ are compensated for by the high density of states near the Fermi level.

\subsection{Current vs. Magnetic Field}

\begin{figure}[h]
\centering
\includegraphics[width=0.45\textwidth]{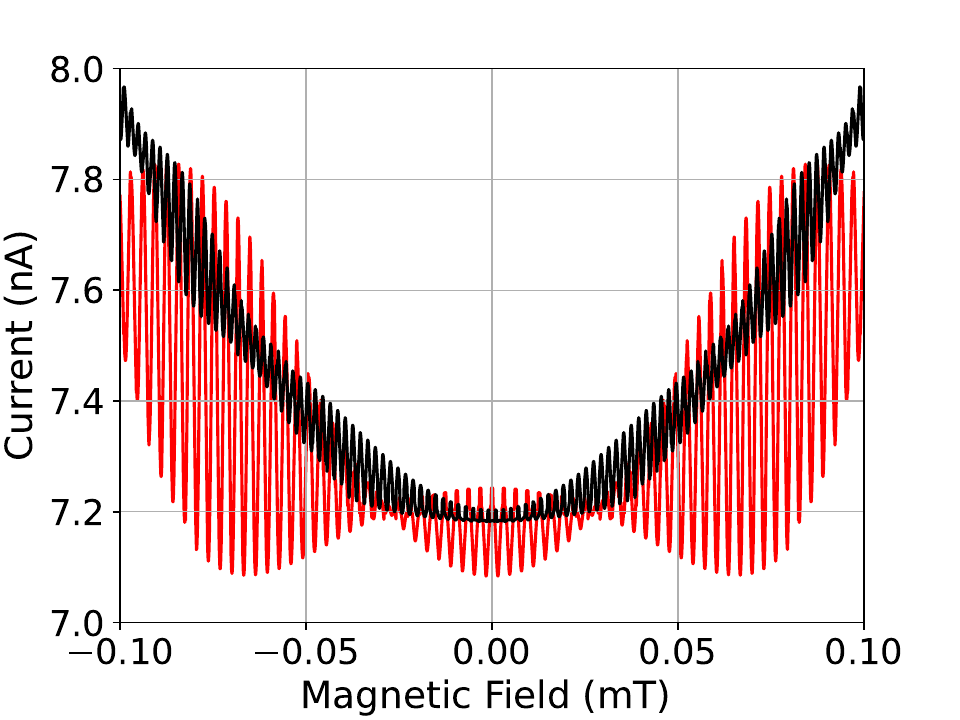}
\caption{(Color online) Plot of the definition (\ref{charge01}) for intermediary voltage ($ 100\hspace{0.05cm}\mu\text{V}$) in terms of the external magnetic field for $\alpha=1$ (dark) and $\alpha=0.8$ (red).} 
\label{medgate}	
\end{figure}

In the intermediary region, we obtain a transient and growing behavior for the current where Ohm's law is not obeyed. In Fig. \ref{currentV}, we show the plot for $I(V,T)$ for $T=40\hspace{0.05cm}\text{mK}$ for different cases of curvature parameter $\alpha$ and zero magnetic field. In Fig. \ref{medgate}, we detail the transient regime for different values of magnetic field when we have $100\hspace{0.05cm}\mu\text{V}$. Comparing it with Fig. \ref{currentmaglowgate}, it is notable how the amplitudes of AB oscillations are suppressed by the voltage around $0\hspace{0.05cm}\text{T}$. 

This influence is also observed in the high-voltage regime, where we also identify in both cases a \textit{staircase} pattern with the increasing magnetic field which confines the oscillations amplitude. Therefore, at a certain point of $B$, the value of $I(V,T)$ will always be higher for $\alpha=1$ compared to $\alpha=0.8$.

\begin{figure}[h]
\centering
\includegraphics[width=0.45\textwidth]{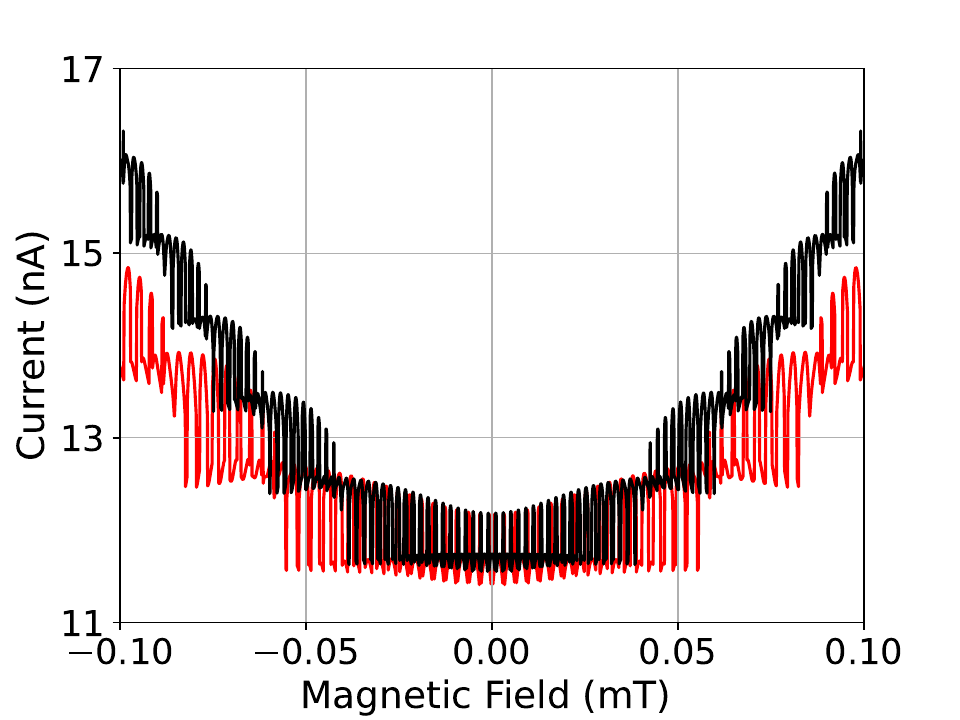}
\caption{(Color online) Plot of the definition (\ref{charge01}) for high voltages ($> 200\hspace{0.05cm}\mu\text{V}$) in terms of the external magnetic field for $\alpha=1$ (dark) and $\alpha=0.8$ (red).}
\label{resonant}	
\end{figure}

\subsection{Current oscillations in $\alpha$}

\begin{figure}[t]
\centering
\includegraphics[width=0.48\textwidth]{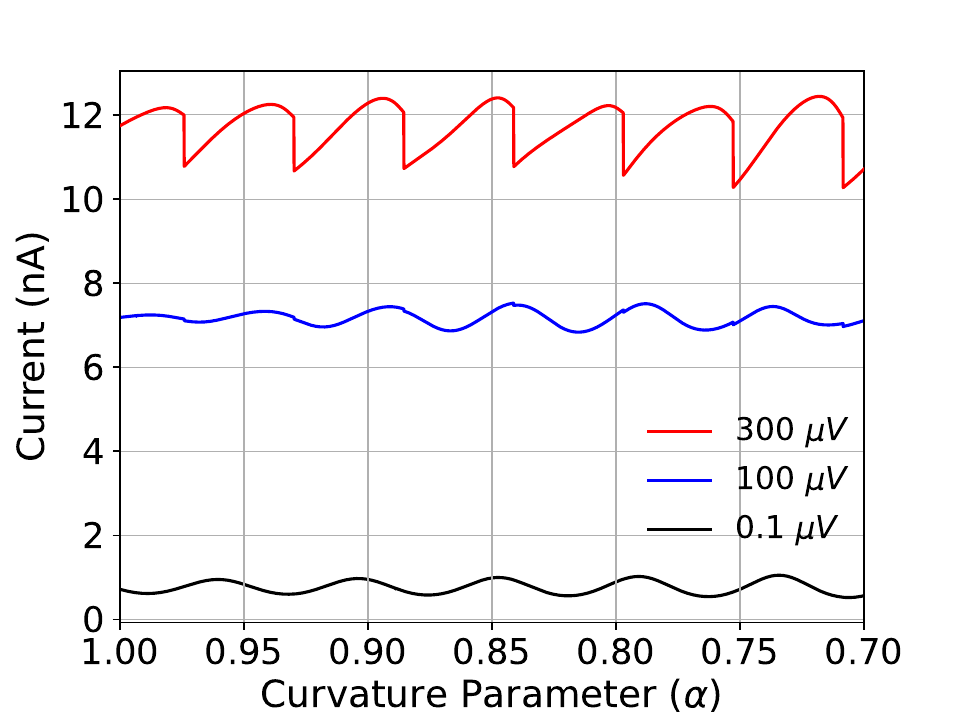}
\caption{(Color online) Plot of the definition (\ref{charge01}) in terms of the curvature parameter $\alpha$ for the low (dark), intermediary (blue) and high-voltage (red) regimes exhibiting distinguish oscillation behaviors. For an intense electric potential, a \textit{sawtooth} pattern is presented.}
\label{currentoscillation}	
\end{figure}

As a last topic of discussion, we would like to do a similar approach applied to magnetoresistance when we fixed the value of the external $B$ field and continuously varied the curvature parameter to find that it oscillates with an almost fixed period when $B=0$ as summarized in TABLE \ref{tab1}. Now, as described in Fig. \ref{currentoscillation}, instead of the magnetic field, we fix the value of the gate voltage for the three cases of interest presented previously.

In the first case (low voltage), we just recover the profile presented in Fig. \ref{oscillationresistance} as a consequence of the inverse relation provided by Ohm's law (\ref{charge05}). The oscillations amplitude increase with the curvature.

For the transient regime, the average value of $I(V,T)$ increases as expected and the sinusoidal oscillation pattern is maintained. However, the amplitudes increase until $\alpha=0.85$ and then reduce until $\alpha=0.7$.

In the high voltage regime, the average value of $I(V,T)$ reaches its maximum as presented in Fig. \ref{currentV}) and we clearly observe a periodic \textit{sawtooth} pattern indicating there are angles of deformation for which the current suddenly decays before starting to increase again.

\section{Conclusions}\label{conclusions}

As we have seen, modifications in the geometry of the device can lead to several implications in the electric charge transport properties. We started by considering the Tan-Inkson model for confining the two-dimensional electron gas. We then let the surface acquire a conical shape where, as immediate consequences, the system had the influence of the background magnetic field attenuated, the eigenenergies suffered a lift due to the factor $m^{2}/\alpha^{2}$ in $L$, and the particles experienced a purely geometrical potential of attractive nature to the ring center. Effects in the energy spectrum were observed as a reduction in the density of states per unity of energy, an increase in the Aharanov-Bohm oscillation period, and a retrenchment in the evolution of the subband bottoms.

In the resonant tunneling process of charged particles between the emitter and the collector acting as two barriers, we argued that the inelastic broadening $\Gamma_{\phi}$ is not affected by the curvature, as the effect on the ring path $2W$ is canceled as the attempt frequency grows. However, since elastic broadenings depend only on $\nu$, the coherent component of transmission increases its impact over the incoherent part.

The downward shift observed in the peaks of the conductance Van-Hove singularities and the increase in both period and amplitude from the oscillating decay with the Fermi level are well described by the effect of curvature in the density of states per unit of energy. No influence of $\alpha$ was observed in the  
distance between the peaks, preserving the experimental measure of the quantity $h\omega_{0}$.

The beating pattern presented by the magnetoresistance oscillations as a result of the interference from the different sets of motion performed by the electrons revealed to be susceptible to variations in the geometry through an increase in both period and amplitude. The dynamics in the envelope functions can be extremely useful, since the average resistance finds its lowest values in the nodes, while the highest ones are located in the antinodes. Their recurrence can be controlled by suitable adjustments in the curvature parameter from intervals of $\Delta\alpha/2$ for $B=0$, presented in TABLE \ref{tab1}.

Another remarkable feature observed was the almost periodic oscillation of magnetoresistance in the variation of the curvature parameter, with a period $\Delta\alpha=0.057$ when there is no external magnetic field. The resistance and conductance peaks are separated by displacements of $\Delta\alpha/2$ (TABLE \ref{tab1}) and their amplitudes continuously increase as a consequence of the effects on both the density of states and elastic broadenings.  For the cases of $B\neq 0$, these oscillations are not periodic due to interference of the states $m<0$ and can present distinct patterns depending on the value of $B$. 

Later, we have investigated how the device performs the electric current through itself upon the application of an electric potential between its terminals. It was identified that for extremely low voltages ($<< 200\hspace{0.05cm}\mu\text{V}$), we still preserve Ohm's law. Furthermore, for high voltages, the system reaches a saturation point where the current acquires stability. It was also observed that in the energetic region between the Fermi level and the bottom of the next subband, the most contributor states are allocated for the total net current. 

Moreover, the charge current dependence with the curvature parameter exhibited a similar periodic pattern observed for the magnetoresistance. However, we  highlighted the \textit{sawtooth} pattern for the high voltage regime, implying the existence of deformation angles for which $I(V,T)$ abruptly decays.

 \section*{Acknowledgments}

The authors would like to express gratitude to the Brazilian agencies CAPES, CNPq, FAPEMA, and the NSF agency. Francisco A. G. de Lira acknowledges the support from the grant CAPES/PDSE 44/2022. Edilberto O. Silva acknowledges the support from the grants CNPq/306308/2022-3, FAPEMA/UNIVERSAL-06395/22, FAPEMA/APP-12256/22 and Coordenação de Aperfeiçoamento de Pessoal de N\'{\i}vel Superior - Brazil (CAPES) - Code 001.  C.D. Santangelo acknowledges support from the NSF grant DMR 2217543. We are also thankful to Manoel M. Ferreira Jr. for useful comments.

\bibliography{references}

\end{document}